\documentclass[
 aip,
 amsmath,amssymb,
 preprint,
]{revtex4-2}

\usepackage[utf8]{inputenc}
\usepackage{amsmath}
\usepackage[dvipsnames]{xcolor}
\usepackage{tabularx}
\usepackage{multirow}

\usepackage{ulem}

\usepackage{tikz}
\usetikzlibrary{arrows.meta}
\tikzset{
  >={Latex[width=2mm,length=2mm]},
            base/.style = {rectangle, rounded corners, draw=black,
                           minimum width=4cm, minimum height=1cm,
                           text centered, font=\sffamily},
            a/.style = { fill=blue!40},
            b/.style = {base, fill=red!40},
            c/.style = { fill=white!40},
            d/.style = {base, minimum width=2.5cm, fill=orange!15,
                           font=\ttfamily},
}

\definecolor{corrections}{rgb}{1.0, 0.0, 0.0}

\definecolor{ykred}{rgb}{1.0, 0.0, 0.0}

\definecolor{jxblue}{rgb}{0.0, 0.0, 1.0}

\definecolor{rygreen}{rgb}{0.0, 0.75, 0.0}

\definecolor{shblue}{rgb}{0.0, 0.7, 1.0}

\definecolor{cmgreen}{rgb}{0.21,0.37,0.23}

\definecolor{ztteal}{rgb}{0.0,0.5,0.5}

\definecolor{vbcolor}{rgb}{0.5,0.5,0.0}

\newcommand{\vect}[1]{\boldsymbol{#1}}

\begin{document}

\title{Nuclear-Electronic Orbital Approach to Quantization of Protons in Periodic Electronic Structure Calculations}
\author{Jianhang Xu*}
\author{Ruiyi Zhou*}
\affiliation{Department of Chemistry, University of North Carolina at Chapel Hill, Chapel Hill, North Carolina 27599, USA}
\author{Zhen Tao}
\author{Christopher Malbon}
\affiliation{Department of Chemistry, Yale University, New Haven, Connecticut 06520, USA}
\author{Volker Blum}
\affiliation{Thomas Lord Department of Mechanical Engineering and Material Science, Duke University, Durham, North Carolina 27708, USA}
\author{Sharon Hammes-Schiffer}
\email{sharon.hammes-schiffer@yale.edu}
\affiliation{Department of Chemistry, Yale University, New Haven, Connecticut 06520, USA}
\author{Yosuke Kanai}
\email{ykanai@unc.edu\\$*$Equal contributions}
\affiliation{Department of Chemistry, University of North Carolina at Chapel Hill, Chapel Hill, North Carolina 27599, USA}

\date{\today}

\begin{abstract} 
The nuclear-electronic orbital (NEO) method is a well-established approach for treating nuclei quantum mechanically in molecular systems beyond the usual Born-Oppenheimer approximation. 
In this work, we present a strategy to implement the NEO method for periodic electronic structure calculations, particularly focused on multicomponent density functional theory (DFT). 
The NEO-DFT method is implemented in an all-electron electronic structure code, FHI-aims, using a combination of analytical and numerical integration techniques as well as a resolution of the identity scheme to enhance computational efficiency.
After validating this implementation, proof-of-concept applications are presented to illustrate the effects of quantized protons on the physical properties of extended systems such as two-dimensional materials and liquid-semiconductor interfaces. Specifically, periodic NEO-DFT calculations are performed for a trans-polyacetylene chain, a hydrogen boride sheet, and a titanium oxide-water interface. 
The zero-point energy effects of the protons, as well as electron-proton correlation, are shown to noticeably impact the density of states and band structures for these systems. These developments provide a foundation for the application of multicomponent DFT to a wide range of other extended condensed matter systems.
\end{abstract}

\maketitle

\section{Introduction}

\par Many physical, chemical, and biological processes depend on quantum mechanical properties of nuclei, such as zero-point energy and hydrogen tunneling effects.
Extensive studies have shown that treating hydrogen atoms quantum mechanically influences both the structural and dynamical properties of water.\cite{PhysRevLett.101.017801,ceriotti_nuclear_2013,ceriotti_nuclear_2016}
Moreover, quantized protons and non-Born-Oppenheimer effects are essential to understanding proton-coupled electron transfer (PCET),\cite{huynh_proton-coupled_2007,weinberg_proton-coupled_2012,hammes-schiffer_proton-coupled_2015,hammes2021integration} which is critical for photosynthesis and respiration as well as many solar cell related processes.
PCET is also a key step in many heterogeneous catalytic processes occurring at solid-liquid interfaces.\cite{li2006ultrafast,wood2013hydrogen,chen2013chemical,barry2021advanced} 
Thus, developing computational methods that incorporate  nuclear quantum effects and non-Born-Oppenheimer behavior for extended condensed phase systems is important for enabling a broad range of applications. 

\par One approach for capturing non-Born-Oppenheimer effects in condensed phase systems is through nonadiabatic dynamics,
where nuclei are propagated on multiple adiabatic potential energy surfaces provided by first principles methods.
Such nonadiabatic dynamics methods include surface hopping,\cite{tully_molecular_1990} Ehrenfest,\cite{li_ab_2005} and multiple spawning.\cite{curchod_ab_2018}
Nuclear quantum effects can be included in simulations using path integral methods,\cite{marx1996ab} which have also been extended to nonadiabatic systems.\cite{chowdhury2021non}
Alternative approaches that treat electrons and nuclei on the same level without invoking the Born-Oppenheimer separation include the exact factorization method\cite{abedi_exact_2010} and multicomponent wave function methods or density functional theory (DFT).\cite{capitani1982non,kreibich_multicomponent_2001,butriy_multicomponent_2007,chakraborty2009properties} 
However, most applications of these types of multicomponent formalisms are restricted to relatively small systems due to the high computational cost.

\par The nuclear-electronic orbital
(NEO)\cite{webb_multiconfigurational_2002,iordanov_vibrational_2003,pak_electron-proton_2004,swalina_explicit_2006,mejia-rodriguez_multicomponent_2019,xu_constrained_2020,xu_full-quantum_2020,pavosevic_multicomponent_2020,hammes-schiffer_nuclearelectronic_2021} method is a well-established multicomponent formalism that is computationally tractable for relatively large molecular systems. 
For many systems, the dominant nuclear quantum effects have been shown to arise from the hydrogen nuclei, i.e., protons.\cite{webb_multiconfigurational_2002,xu_full-quantum_2020}
Within the NEO framework, specified protons are treated quantum mechanically on the same level as the electrons, while the remaining nuclei are treated classically. In some cases, other nuclei can also be treated quantum mechanically within this framework. 
The simplest starting point of the NEO method is NEO Hartree-Fock (NEO-HF) theory,\cite{webb_multiconfigurational_2002}
where the nuclear-electronic wave function is described by a single product of nuclear and electronic Slater determinants.
This approximation fails to capture the correlation between electrons and protons, however, leading to a poor description of nuclear quantum effects. 
One approach to systematically improve the description of electron-proton correlation is through post Hartree-Fock wave function approaches, such as perturbation theory methods,\cite{swalina2005alternative,pavosevic2020multicomponent} coupled-cluster methods,\cite{pavosevic2018multicomponent} and multireference methods.\cite{webb2002multiconfigurational,fajen2021multicomponent}
However, the high order polynomial scaling cost of wave function methods limits their applicability to periodic or extended condensed matter systems.

\par A more computationally practical approach is NEO-DFT,\cite{pak_density_2007,chakraborty_development_2008} 
where the electron-proton correlation is approximated as a functional of the electron and proton densities.
In recent years, different levels of electron-proton correlation functionals have been introduced and successfully used to describe electron-proton coupling, including local density approximation (LDA) functionals, such as  epc17 \cite{yang_development_2017,brorsen_multicomponent_2017} and epc18, \cite{brorsen_alternative_2018} and generalized gradient approximation (GGA) functionals, such as epc19. \cite{tao_multicomponent_2019}
According to the multicomponent Kohn-Sham formalism, the non-interacting reference system is represented as a product of electronic and nuclear Slater determinants. Within the NEO framework, the electronic and protonic Kohn-Sham equations are solved iteratively to determine the total energy of the system and the corresponding electron and proton densities in the field of the classical nuclei. Given the close analogy of this approach to conventional electronic DFT, NEO-DFT is particularly robust and applicable to extended systems.
The goal of this work is to extend multicomponent NEO-DFT to calculations of periodic systems.

\par The remainder of this paper is organized as follows.
In Sec. \ref{sec:method}, we provide a brief introduction to the multicomponent NEO-DFT framework and demonstrate a computationally convenient way to implement NEO-DFT in periodic electronic structure codes. Specifically,
a density fitting (i.e., resolution of the identity) approach is used to describe the electrostatic interactions and exchange effects of the quantized protons. 
In Sec. \ref{sec:validation}, we validate our  implementation of periodic NEO-DFT to the FHI-aims code \cite{blum_ab_2009} by comparing results for an isolated system to results obtained with existing Q-Chem code. \cite{epifanovsky2021software} 
In Sec. \ref{sec:example}, we present three proof-of-principle examples of applications to extended systems to showcase the capabilities of the new periodic NEO-DFT method. These applications are a trans-polyacetylene chain, a hydrogen boride sheet, and water molecules at a titanium oxide surface.
In Sec. \ref{sec:conclusion}, we conclude our work with a brief summary and discussion of future directions.

\section{Theory and Computational Methods}
\label{sec:method}
\subsection{Multicomponent DFT via NEO method for Kohn-Sham system}

The NEO formalism in the framework of multicomponent DFT leads to a set of coupled electron-proton Kohn-Sham (KS) equations,

\begin{align}
    \hat{H}^e_{\bf{k}} \psi^e_{i,\bf{k}} (\vect r^e) &= \left[-\frac{1}{2}\nabla^2+v_{\text{eff}}^e(\vect r^e)\right] \psi^e_{i,\bf{k}} = \epsilon^e_{i,\bf{k}} \psi^e_{i,\bf{k}} (\vect r^e), \label{eq:KS_e}\\
    \hat{H}^p \psi^p_i (\vect r^p)&= \left[-\frac{1}{2M^p}\nabla^2+v_{\text{eff}}^p(\vect r^p)\right] \psi^p_i = \epsilon^p_i \psi^p_i (\vect r^p), \label{eq:KS_p}
\end{align}
where $M^p$ is the proton mass, $\psi^e_{i,\bf{k}}$ and $\epsilon^e_{i,\bf{k}}$ are the KS orbitals and eigenvalues of the electrons, and $\psi^p_{i}$ and $\epsilon^p_{i}$ are the KS orbitals and eigenvalues of the protons. 
The effective potentials are defined as 
\begin{align}
    v_{\text{eff}}^e(\vect r^e) &= v_{\text{ext}}(\vect r^e) + v_{\text{es}}^{e}(\vect r^e)-v_{\text{es}}^{p}(\vect r^e)+\frac{\delta E_{\text{xc}}^{e}[\rho ^e]}{\delta \rho^e}+\frac{\delta E_{\text{epc}}[\rho^e,\rho^p]}{\delta \rho^e}, \label{equ:v_eff_e}\\
    v_{\text{eff}}^p(\vect r^p) &= -v_{\text{ext}}(\vect r^p)-v_{\text{es}}^{e}(\vect r^p)+v_{\text{es}}^{p}(\vect r^p)+\frac{\delta E_{\text{xc}}^{p}[\rho ^p]}{\delta \rho^p}+\frac{\delta E_{\text{epc}}[\rho^e,\rho^p]}{\delta \rho^p}.\label{equ:v_eff_p}
\end{align}
where $v_{\text{ext}}$, $v_{\text{es}}^{e}$, and  $v_{\text{es}}^{p}$ are electrostatic potentials for the classical nuclei, electrons, and quantum protons. 
$E_{\text{xc}}^e$ and $E_{\text{xc}}^p$ are the exchange-correlation energies of the electrons and the quantum protons, respectively. 
$E_{\text{epc}}$ is the correlation energy between the electrons and quantum protons, 
which is given as a functional of both the electron density and the proton density in the multicomponent DFT formalism.\cite{chakraborty_development_2008,brorsen_multicomponent_2017,yang_development_2017,brorsen_alternative_2018,tao_multicomponent_2019} 
In most cases, the KS wave functions for the quantum protons are extremely localized in real space such that the overlap of proton wave functions from different unit cells is close to zero.
Thus, the Brillouin zone (BZ) integration can be neglected for the protons even for extended systems, whereas the electron KS wave functions require the BZ integration.  
Further numerical evidence for the validity of this $\Gamma$-point approximation for protons is discussed in Sec.\ref{subsec:C2H2}.
With Equations \ref{eq:KS_e} and \ref{eq:KS_p}, the ground state of the multicomponent system of electrons and quantum protons needs to be solved self-consistently in the presence of the electrostatic potential from the classical nuclei.

\par In the following subsections, we present a convenient framework for implementing the NEO method in periodic electronic DFT codes, 
thus expanding the capability of the NEO approach to study condensed matter systems. Our approach here is to make the periodic NEO implementation as modular as possible with minimal modifications to the underlying electronic structure code. 
The main idea is to separate the Hamiltonian for protons and electrons into two parts: 
uncoupled terms ($H^p_p$, $H^e_e$) and coupled terms ($H^p_{pe}$, $H^e_{ep}$),
\begin{align}
    &H^p_p \equiv -\frac{1}{2M^p}\nabla^2 - v_{\text{ext}}(\vect r^p) + v_{\text{es}}^{p}(\vect r^p) + \frac{\delta E_{\text{xc}}^{p}[\rho ^p]}{\delta \rho^p}, \\
    &H^e_e \equiv -\frac{1}{2}\nabla^2 + v_{\text{ext}}(\vect r^e) + v_{\text{es}}^{e}(\vect r^e) + \frac{\delta E_{\text{xc}}^{e}[\rho ^e]}{\delta \rho^e},\\
    &H^p_{pe} \equiv -v_{\text{es}}^{e}(\vect r^p) + \frac{\delta E_{\text{epc}}[\rho^e,\rho^p]}{\delta \rho^p},\\
    &H^e_{ep} \equiv - v_{\text{es}}^{p}(\vect r^e) + \frac{\delta E_{\text{epc}}[\rho^e,\rho^p]}{\delta \rho^e}.
\end{align}

To solve the proton KS equation (Eq. \ref{eq:KS_p}), the uncoupled $H^p_p$ terms are constructed within the NEO approach, 
and the coupled $H^p_{pe}$ terms are obtained from the ground state electronic structure via the periodic DFT calculation.
Then the coupled term for the electron KS equation (Eq. \ref{eq:KS_e}),  $H^e_{ep}$, is calculated via the NEO method and passed back to the electronic DFT calculation as a correction term for solving the electron KS equation.
The NEO methods have been implemented using tailored Gaussian-type orbitals as the basis set for the protons,\cite{webb_multiconfigurational_2002,yu_development_2020} whereas most periodic DFT codes are based on other types of basis sets. 
Thus, decoupling of the electron and proton Hamiltonians requires careful considerations to be discussed in the following subsections.
In the present work, numerically tabulated atom-centered orbitals (NAOs), as implemented in the FHI-aims code, are used for the electronic basis sets.
The numerical framework, however, can be implemented with a wide range of periodic DFT electronic structure codes. 

\subsection{Basis sets and atomic integrals for electrons and protons}
\par The electronic wave functions are expressed as linear combinations of grid-based NAOs in the FHI-aims code. 
The use of such an NAO basis set for studying extended systems has been discussed in Ref. \citenum{blum_ab_2009}. 
The more accurate description of the sharp electron peak close to the nucleus is a distinct advantage of the NAO approach over other types of basis sets. 
For quantum protons, the cusp at the center is less steep, and Gaussian type orbital (GTO) basis sets are chosen to describe the proton wave functions. Thus, the orbitals can be expressed as
\begin{align}
    \psi^{e}_{i,\bf{k}}(\vect r^e)&=
    \sum_{\vect{N}}\sum_{\mu}c_{\mu i \bf{k}}e^{i\bf{k}\cdot \vect{T(N)}}\phi^{e:\text{NAO}}_{\mu,\vect N}(\vect r^e),\\
    \psi^{p}_i(\vect r^p)&=
    \sum_{\vect{N}}\sum_{m}c_{m i}\phi^{p:\text{GTO}}_{m,\vect N}\left(\vect r^p\right), \label{eq:pwavefunction}
\end{align}  
and
\begin{equation}
\begin{aligned}
    \phi^{e:\text{NAO}}_{\mu,\vect N}(\vect r^e) & = \phi^{e:\text{NAO}}_{\mu}(\vect r^e-\vect{R}_{\mu}+\vect T(\vect N)), \\
    \phi^{p:\text{GTO}}_{m,\vect N}\left(\vect r^p\right) &= \phi^{p:\text{GTO}}_m\left(\vect r^p-\vect {R}_{m}+\vect T(\vect N)\right),
\end{aligned}
\end{equation}
where $\vect N= (N_1,N_2,N_3)$ are the neighboring unit cells, $\vect T(\vect M)$ is a translation vector of cell $\vect M$, and $\vect R_{x}$ is the coordinate of the center of basis function $x$.
In the mixed usage of NAO and GTO basis sets within this multicomponent DFT formalism, it is convenient to employ two different types of integration strategies: analytical integrals for the GTOs and numerical real-space grid integration. 
For constructing the proton Hamiltonian, analytical integrals are adopted by calling the Libcint library \cite{10.1002/jcc.23981} when only GTO basis functions are involved.
In other cases, numerical integrals are performed by mapping proton GTO basis functions and potentials onto the real-space grid points, which are defined via the FHI-aims code.\cite{havu_efficient_2009}

\par 
Because protons are highly localized, the wave functions of the quantum protons are treated within the $\Gamma$-point approximation to the BZ integration but with periodic boundary conditions (PBC). 
Similarly to the case of electrons,\cite{blum_ab_2009} the proton Hamiltonian of the PBC system in the GTO basis can be expressed as
\begin{equation}
h_{m n}^p=\sum_{\vect M, \vect N}\left\langle\phi_{m, \vect M}^p|\hat{h}| \phi_{n, \vect N}^p\right\rangle=\sum_{\vect M, \vect N} h_{m, n}^p(\vect M, \vect N).
\label{h_pbc}
\end{equation}
Equation \ref{h_pbc} diverges if it is used directly, summing over both $\vect M$ and $\vect N$ cells, because it only depends on $\vect N-\vect M$.
For analytical integrals, we sum over $\vect N-\vect M$ to avoid divergence, 
\begin{equation}
    h_{m n}^{p,\text{analytical}}=\sum_{\vect N-\vect M}\left\langle\phi^p_{m, 0}|\hat{h}| \phi^p_{n, \vect N-\vect M}\right\rangle.
    \label{h_analytical}
\end{equation}
On the other hand, the numerical grid integrals are calculated over only grid points in the center simulation cell (0-cell),\cite{knuth_all-electron_2015} 
\begin{equation}
    h_{m n}^{p,\text{numerical}}=\sum_{\vect M, \vect N}\int_\text{unit cell}\phi^p_{m, \vect M}(\vect r)h(\vect r)\phi^p_{n, \vect N}(\vect r)\text d \vect r.
\end{equation}
As illustrated in Figure 1 of Ref. \citenum{knuth_all-electron_2015}, the full integral over the volume in which the two basis functions in Eq. (\ref{h_analytical}) are zero can then be recovered by adding partial integrals $h_{m n}^{p,\text{numerical}}$ that extend only over a single unit cell.
Therefore, only the basis functions that are in the immediate vicinity of the 0-cell are required in the calculation, and the divergence in Eq. \ref{h_pbc} is avoided.
In practice, we only include the image basis functions if their value on any grid point in the 0-cell exceeds a given threshold.
The charges of image proton atoms are included within a preset radius cutoff determined by the distances at which their screened electrostatic potentials influence the 0-cell.

\subsection{Proton Hamiltonian}
\label{sec:proton_h}
In addition to the kinetic energy of the quantum protons, $\hat{T}^{p}=-\frac{1}{2M^p}\nabla^2$, that is constructed by analytical integrals,
the rest of the Hamiltonian of the quantum protons is defined with the effective potential energy operator given by Eq. \ref{equ:v_eff_p}.
The effective potential contains several terms. 
The electrostatic interaction between electrons and quantum protons can be calculated with numerical integrals by reading the electrostatic potential of the electrons, $v_{\text{es}}^{e}(\vect r^p)$, on the grid points.
The electrostatic potential from the quantum protons, $v_{\text{es}}^{p}(\vect r^p)$, is obtained separately as discussed in detail in Sec.\ref{sub:electro}. 
Furthermore, in DFT calculations for periodic systems,
the electrostatic potential contributed by electrons, $v_{\text{es}}^{e}(\vect r^p)$, and all nuclei, $v_\text{ext}^\text{DFT}(\vect r^p) = \sum_{I}\frac{Z^I}{|\vect r-\vect r^I|}$, is evaluated on the grid points, and the sum of these two terms is stored together as a Hartree potential to avoid a  singularity. 
Thus, $v_\text{ext}(\vect r^p)$, defined as the external potential due to only the classical nuclei, is obtained by subtracting the quantum proton contribution from $v_\text{ext}^\text{DFT}(\vect r^p)$.
The remaining terms in the proton Hamiltonian are exchange-correlation effects for the quantum protons and electron-proton correlation effects. 
These two terms are discussed in Sec.\ref{sub:exchange} and Sec.\ref{sub:epc}, respectively. 
In this section, we omit the superscript $p$ from all proton basis functions $\phi^p$ and proton coordinates $\vect r^p$ for brevity where it is obvious. 

\subsubsection{Resolution of the identity}
\label{subsubsec:ri}
\par 
Before discussing the above mentioned terms in the following subsections, we briefly address the use of the resolution of the identity (RI) scheme\cite{ren_resolution--identity_2012} in our work. 
The electrostatic potential term $v_{\text{es}}$ appears in both the proton and electron Hamiltonians (Eqs. \ref{eq:KS_e} and \ref{eq:KS_p}).
It is evaluated by numerical integrals and stored on grid points, so that it can be used for updating the electron Hamiltonian later.
To evaluate the proton electrostatic potential, the proton density $\rho^p$ is expanded using an auxiliary Gaussian basis set $\Phi_{\text{aux}}$:
\begin{equation}
\label{rho_gau}
\begin{aligned}
        \rho^p(\vect r) 
        &= \sum_{m,n}D_{mn}\phi_m(\vect r)\phi_n(\vect r)= \sum_{\mu} c_{\mu} \Phi_{\mu}(\vect r) \\
        & = \sum_{\mu} c_{\mu}N_\mu r^{l_\mu} e^{-\alpha_\mu r^2} Y_{l_\mu m_\mu}(\theta,\phi),
\end{aligned}
\end{equation}
where $D_{mn}=\sum_{i}c_{mi}^{*}c_{ni}$ is the density matrix for the protons, and $N_\mu$, $l_\mu$, and $\alpha_\mu$ are the normalization coefficient, angular momentum, and Gaussian exponent, respectively, of the auxiliary basis function $\mu$.
The fitting coefficient $c_{\mu}$ is calculated using the Coulomb metric as follows,
\begin{align}
    c_{\mu} &= \sum_{m,n}D_{mn}C_{mn}^\mu,\\
    C_{mn}^\mu &= \sum_\nu (mn|\nu)(\nu|\mu)^{-1},
\end{align}
where 
\[(mn|\nu)=\int \phi_m(\vect r)\phi_n(\vect r) \frac{1}{|\vect r - \vect r'|} \Phi_\nu(\vect r') \text d\vect r\text d\vect r', \] 
and 
\[(\nu|\mu)=\int \Phi_\nu(\vect r) \frac{1}{|\vect r - \vect r'|} \Phi_\mu(\vect r') \text d\vect r\text d\vect r'.\]
For molecular systems, this approach has been widely employed to speed up the calculation of Coulombic and exchange interactions in large systems by reducing the scaling cost of 4-index atomic integrals.\cite{mejia-rodriguez_multicomponent_2019,pavosevic_multicomponent_2021} 

\par  
For extended systems, a localized version of RI (RI-LVL)\cite{ihrig_accurate_2015,levchenko_hybrid_2015} is employed to enhance numerical efficiency. 
The RI-LVL is a localized RI scheme implemented in FHI-aims by \citeauthor{ihrig_accurate_2015}\cite{ihrig_accurate_2015}, and the technique exploits the form of the local matrix (L) and the Coulomb matrices (V) used to compute the two-electron integrals.
In this scheme, the fitting coefficient $C_{mn}^\mu$ is calculated similarly as for RI if the center of the auxiliary basis function $\Phi_\mu^{\bf K}$ is the same as the corresponding center of the proton basis function $\phi_m^{\bf I}$ or $\phi_n^{\bf J}$
and is set to zero otherwise. 
In this case,
\begin{equation}
     C_{mn}^\mu = 
     \begin{cases}
        \sum_\nu (mn|\nu)L^{\bf{IJ}} & \text{if } \bf K=\bf I\text{ or } \bf J\\
        0 & \text{otherwise},
     \end{cases}
     \label{eq:CRILVL}
\end{equation} 
where the indices $\bf I,\bf J,\bf K$ denote atom centers of the basis functions, and $L^{\bf{IJ}} =[(\nu|\mu)^{\bf{IJ}}]^{-1}$ with the inverse calculated only for auxiliary basis functions $\Phi_\nu$ and $\Phi_\mu$ centered on $\bf I$ and $\bf J$, respectively. 
It essentially amounts to approximating the orbital product of two distinct atomic centers $\bf I,J$ by fitting to the auxiliary basis at these two centers.\cite{ihrig_accurate_2015} 
The RI-LVL scheme is highly useful for our implementation here for two main reasons. 
First, it avoids fitting of long-range Coulomb interactions between spatially distant basis functions in different PBC cells, although it requires higher angular momentum of the auxiliary basis, e.g., $l$ = 6 or 8 in Table 12 of Ref. \citenum{ihrig_accurate_2015}
to retain sufficient accuracy for electron system. 
Second, it allows us to exploit the localized nature of the proton basis, offering numerical advantages over other mixed plane-wave density fitting schemes or long-range Coulomb integral implementations for quantum protons. 
Convergence tests of this RI-LVL scheme are provided in the Supplementary Material. 

\subsubsection{Electrostatic potential via Gaussian multipoles}
\label{sub:electro}
The electrostatic potential from the electrons, $v_\text{es}^e(\vect r)$, in Eq. \ref{equ:v_eff_p} is obtained from the Hartree potential, which is readily available in the existing DFT implementation within the FHI-aims code.
We could also use existing subroutines in the FHI-aims code for calculating the electrostatic potential from the protons,  $v_\text{es}^p(\vect r)$. However, we pursue a different approach here to reduce the computational cost and to make the NEO implementation as modular as possible to facilitate its transferability to other  electronic structure codes. 
In our approach, $v_\text{es}^p(\vect r)$ is solved analytically using the Green's function of the Poisson equation.
In spherical coordinates, it takes the form
\begin{equation}
\label{eq:vgreen}
    U(\vect r)= \sum_{l=0}^{l_\text{max}} \sum_{m=-l}^{l} \frac{4\pi}{2l+1}[r^lp_{l,m}(r)+\frac{q_{l,m}(r)}{r^{l+1}}]Y_{l,m}(\theta,\phi),
\end{equation}
where
\begin{align*}
    p_{l,m}(r)&=\int_r^\infty \int \frac{1}{r'^{l+1}}\rho(r',\theta,\phi)Y_{lm}r'^2\text d\Omega_{lm}\text dr' \\
    q_{l,m}(r)&= \int_0^{r} \int r'^{l}\rho(r',\theta,\phi)Y_{lm}r'^2\text d\Omega_{lm}\text dr'.
\end{align*}
Because the density of the quantum protons is expressed as the sum of Gaussian multipoles for RI-LVL, $v_\text{es}^p(\vect r)$ is calculated as the sum of the electrostatic potentials of all the Gaussian multipoles. In particular,
\begin{equation} 
\label{eq:vespr}
    v_\text{es}^p(\vect r) = \sum_\mu U_\mu(\vect r - \vect r_{\bf I_\mu}),
\end{equation}
where $\bf I_\mu$ denotes the center of the density fitting proton basis function $\mu$, $\vect r_{\vect I_\mu}$ denotes its coordinates, and
\begin{align}
\label{eq:u_sr}
     U_\mu(\vect r) &= \frac{4\pi}{2l+1}[r^{l_\mu} p_\mu(r)+\frac{q_\mu(r)}{r^{{l_\mu}+1}}]Y_{{l_\mu},m_\mu}(\theta,\phi), \\
    p_\mu(r)&= c_{\mu}N_\mu \frac{1}{2\alpha_\mu} e^{-\alpha_\mu r^2}, \\
    q_\mu(r)&= c_{\mu}N_\mu \int_0^{r} r'^{2l_\mu+2}e^{-\alpha_\mu r'^2}\text dr'.
\end{align}
A detailed derivation of this electrostatic potential from Gaussian multipoles is given in Appendix \ref{app:gau}. 

\par 
At sufficiently large $r$, $p_\mu(r)$ vanishes and $q_\mu(r)$ converges to a constant value, and Eq. \ref{eq:u_sr} reduces to
\begin{equation}
     \label{eq:u_lr}
     U_\mu(\vect r) =\frac{4\pi}{2l+1}\frac{C_{\mu}}{r^{l+1}}Y_{l,m}(\theta,\phi),
\end{equation}
where 
$C_{\mu} = c_\mu N_\mu\frac{\sqrt{\pi}}{2\sqrt{\alpha}}\frac{(2l+1)!!}{(2\alpha)^{l+1}}$ 
represents a classical spherical multipole that has no $r$ dependence.
This simplification enables us to calculate and sum up the classical multipoles for all density fitting basis functions with the same basis function center $\bf I_\mu$. 
The only distance dependent term in the electrostatic potential is the $1/r^{l+1}$ term, 
which can be calculated, for example, using the Ewald summation method\cite{delley_fast_1996} for each proton center in periodical cases.

\par 
Alternatively, considering that the electrostatic potential contributed by the classical point charges of hydrogen nuclei ($v_{1/r}^p$) is already used by conventional DFT,
one can calculate only the difference between $v_{1/r}^p$ and the electrostatic potential of the quantum protons ($v_\text{es}^p$), as given by Eqs. \ref{eq:u_sr} - \ref{eq:u_lr}.
This difference is defined as a correction potential $v_\text{es-corr}^p = v_\text{es}^p - v_{1/r}^p$, which is then used to update the total electrostatic potential with the contribution from the quantum protons. 
For ground state NEO-DFT calculations in which the protons are well-localized in real space around the basis centers,
$v_\text{es-corr}^p$ can be calculated as the sum of the contributions from individual proton basis function centers:
\begin{equation}
    v_\text{es-corr}^p=\sum_{\bf I} v_{\text{es-corr},{\bf I}}^p =\sum_{\bf I} (v_\text{es,{\bf I}}^p - v_{1/r,{\bf I}}^p) = \sum_{\bf I}( \sum_{\mu \in {\bf I}}U_\mu(\vect r - \vect r_{\bf I_\mu}) - \frac{1}{|\vect r - \vect r_{\bf I}|}).
\end{equation}
For each proton basis function center ${\bf I}$,
$v_{\text{es-corr},{\bf I}}^p$ can be approximated as a short-range potential for the following reason. 
Around each proton basis function center ${\bf I}$, a local domain $S_{\bf I}$ can be found such that the proton density is normalized, i.e., $ \int_{S_{\bf I}}\rho^p(\vect r) \text d\vect r=1 $. 
We numerically validate the existence of such a local domain $S_{\bf I}$ by calculating the sum of the local Gaussian charges, 
namely the charge distribution from the Gaussian auxiliary basis for $l=0$. Finding the value for this quantity to be one, where
\begin{equation}
    \sum_{\mu \in {\bf I}}c_{\mu}N_{\mu}\int_{S_{\bf I}} 
    e^{-\alpha_\mu r^2}r^{l_\mu} Y_{l_\mu,m_\mu}(\theta,\psi)d\vect r =  \sum_{\mu \in {\bf I}} \frac{\sqrt{32\pi}}{N_\mu} c_\mu \delta_{l_\mu,0} = 1,
\end{equation}
indicates the presence of such a local domain within the auxiliary basis, and in practice, our calculations yield the value of one to within 10$^{-5}$ for all cases considered in this work.

Therefore, the electrostatic potential of these Gaussian charges cancels out the potential generated from a point charge at long range and yields a short-range error function in the leading term,
\begin{equation}
    v_{\text{es-corr},\bf{I}}^p(\vect r)|_{l=0}\sim\sum_{\alpha} c_{\alpha}\frac{erfc(\alpha r)}{r},
\end{equation}
where $c_{\alpha}$ is the proton density fitting coefficient for the Gaussian multipole with the exponent $\alpha$ and $l=0,m=0$ (see Section \ref{subsubsec:ri}).
The remaining parts of this correction potential (i.e., $l\geq 1$) are treated as classical multipoles, which decay as $c_{\alpha}\cdot r^{-(l+1)}$ with increasing radius. 
Additionally, their fitting coefficients $c_{\alpha}$ are at least one order of magnitude smaller in comparison to the $l=0$ term. 

Thus, the correction potential $v_\text{es-corr}^p$ can be treated as a short-range potential and only  calculated within a given cutoff radius $r_\text{es}^\text{cut}$ with respect to the proton basis function center.
A numerical assessment of how calculations converge using this short-range correction potential $v_\text{es-corr}^p$ with respect to the cutoff radius is given in Section \ref{sec:example}. This newly-proposed scheme for evaluating the electrostatic potential for protons 
by constructing the short-ranged correction potential is computationally attractive, particularly for cases in which protons are fairly well localized. 

\subsubsection{Exchange correlation energy for quantum protons}
\label{sub:exchange}

\par 
Because of the spatially localized nature of protons, the exchange and correlation effects among protons are negligibly small when compared to the electron-electron and electron-proton counterparts.\cite{chakraborty_development_2008,auer2010localized} 
A quantitative analysis in Ref. \onlinecite{pavosevic_multicomponent_2020} indicates that the proton-proton exchange and correlation are often 6 to 10 orders of magnitude smaller than their electron-electron counterparts.
Thus, in practice, $E_{\text{xc}}^{p}[\rho ^p]$ in Eq. \ref{equ:v_eff_p} is replaced by the Hartree-Fock exact exchange $K^{p}$ for the quantum protons. 
The exchange energy is calculated from
\begin{equation}
    K^{p}_{mn}=\sum_{k,l}D_{kl}(mk|nl),
\end{equation}
where $D_{kl}$ is the density matrix and the two electron integral is 
\begin{equation}
(mk|nl) = \int \phi_m(\vect r)\phi_k(\vect r) \frac{1}{|\vect r - \vect r'|} \phi_n(\vect r')\phi_l(\vect r') \text d\vect r\text d\vect r'.
\end{equation}
Within the RI-LVL approach, the integral is calculated as
\begin{equation}
\label{mknl}
    (mk|nl) = \sum_{\mu,\nu} (mk|\mu)(\mu|\nu)^{-1}(\nu|nl) =\sum_{\mu,\nu} C^\mu_{mk}(\mu|\nu)C^\nu_{nl},
\end{equation}
where $C^\mu_{mk}$ and $C^\nu_{nl}$ are the RI-LVL fitting coefficient in Eq. \ref{eq:CRILVL}.
Although proton-proton exchange terms are typically negligible, \cite{auer2010localized,pavosevic_multicomponent_2020} the diagonal exchange terms are required to counteract the self-interaction.

For periodic systems, the $\Gamma$-point approximation for quantum protons limits the calculation of $K^p_{mn}$ to only basis functions $m$ and $n$ within the same simulation cell (i.e., the 0-cell).\cite{tymczak_linear_2005} 
For protons near the boundary, we introduce a modified minimum image convention (MIC)\cite{tymczak_linear_2005,irmler_robust_2018} to ensure the translational invariance of $K^p$.
With MIC, the auxiliary basis Coulomb terms $(\mu|\nu)$ in Equation \ref{mknl} are constructed from the nearest images of $\Phi_\mu$ and $\Phi_\nu$.
Protons are highly localized in the ground state, and thus exchange effects between different protons is essentially negligible. 
Thus, only the exchange effect generated from the same proton, the so-called self-interaction, is included in the Hamiltonian.
Previous work has shown that this NEO Hartree approximation with the self-interaction correction is sufficient to give accurate NEO-DFT results as compared to the full NEO Hartree-Fock results.\cite{auer2010localized,xu_full-quantum_2020} 
Note that, in our implementation, the exchange effects over multiple basis function centers can be included in a straightforward manner by controlling a radius cutoff parameter between centers. 
This might be convenient for future real-time simulations of proton transfer reactions and excited state dynamics in which protons are more delocalized.

\subsubsection{Electron-proton correlation functional}
\label{sub:epc}

\par In order to evaluate the electron-proton correlation $E_{epc}[\rho^p,\rho^e]$ in Eq. \ref{equ:v_eff_p}, we employ the epc17-2 electron-proton correlation functional,\cite{yang_development_2017,brorsen_multicomponent_2017} although other approximations can be readily employed. 
The epc17-2 functional is a local density approximation to the density-based correlation functional of electrons and protons in the context of multicomponent DFT.
This functional is defined as
\begin{equation}
\label{epc17}
    E^c_{epc17}[\rho^p,\rho^e]=-\int \text d\vect R \frac{\rho^p(\vect R)\rho^e(\vect R)}{a-b\rho^p(\vect R)^{\frac{1}{2}}\rho^e(\vect R)^{\frac{1}{2}}+c\rho^p(\vect R)\rho^e(\vect R)},
\end{equation}
where $a,b,c$ are parameters defined as $a=2.35$, $b=2.4$, 
and $c=6.6$.
The electron-proton correlation potential is then given by $v_\text{epc}^x = {\delta E_{epc}}/{\delta \rho^x}$, 
where $x$ denotes either electron or quantum proton.
For evaluating the correlation functional given in Eq. \ref{epc17} and its corresponding potential, both the electron and quantum proton densities are represented on the grid points, and the integral is computed numerically.

\subsection{Electron Hamiltonian}
\label{sec:electron_h}

\par Our approach is to make our periodic NEO implementation as independent as possible from the underlying electronic structure code (i.e., FHI-aims). The electron Hamiltonian requires minimal modifications with the effective potential, given by Eq. \ref{equ:v_eff_e}. For solving the electron NEO-KS equation, the  electron-proton coupled Hamiltonian terms in Eq. \ref{equ:v_eff_e} are treated as additional correction terms to the regular KS equation for electrons.
The electrostatic potential contributed by the quantum protons $v^p_\text{es}$ and the electron-proton correlation potential $v_\text{epc}^e$ are tabulated on the grid points, as mentioned in the previous section. 
In actual calculations, $v_\text{epc}^e$ is simply combined with the exchange-correlation potential for the electrons, $v_\text{xc}^e$, and the eigenvalue problem for the electron Hamiltonian is solved within existing electronic structure routines.

\subsection{NEO-DFT total energy}
\par In practical DFT calculations, the total energy is most conveniently obtained as
\begin{equation}
    E_\text{tot}=\sum_{l=1}f_l\epsilon_l-\int v_{xc}^e(\vect r)\rho^e(\vect r^e)\text d\vect r^e+E_{xc}^e[\rho^e(\vect r^e)]-\frac{1}{2}\int v_\text{es}^e{(\vect r^e)}\rho^e(\vect r^e)\text d\vect r^e+E_\text{nuc-nuc},
\end{equation}
where $f_l$ and $\epsilon_l$ are the occupation number and KS eigenvalues for the state $l$. 
The second and third terms account for the exchange-correlation energy. 
The fourth term cancels out the double counting of electron-electron electrostatic interactions in the eigenvalue summation.
The last term represents the classical nuclear-nuclear interactions. 

In the context of multicomponent DFT, the NEO-DFT total energy is 
\begin{equation}
\begin{aligned}
     E^\text{NEO}_\text{tot}=
     &\sum_{l=1}f_l\epsilon_l^\text{NEO}-\int v_{xc}^{e-\text{NEO}}(\vect r^e)\rho^e(\vect r^e)\text d\vect r^e+E_{xc}^{e-\text{NEO}}[\rho^e(\vect r^e)]-\\ &\frac{1}{2}\int v_\text{es}^{e-\text{NEO}}{(\vect r^e)}\rho^e(\vect r^e)\text d\vect r^e +E_\text{nuc-nuc}+\Delta E^\text{NEO}_\text{nuc-nuc}.
\end{aligned}
\end{equation}
Here, $\epsilon_l^{\text{NEO}}$ are the KS eigenvalues from the NEO electron Hamiltonian. Moreover, 
$v_{xc}^{e-\text{NEO}}$ and $E_{xc}^{e-\text{NEO}}$ are the exchange-correlation potential and energy for the NEO electron Hamiltonian,
and they include electron-proton correlation as well as electron-electron exchange-correlation.
The form of the double counting electrostatic term is unchanged.
The last term is responsible for the quantum proton contributions and can be expressed as
\begin{equation}
    \Delta E^\text{NEO}_\text{nuc-nuc}=T^p+J^p-K^p + \Delta E^\text{NEO}_{\text{nuc},0},
\end{equation}
where $T^p,J^p$ and $K^p$ are the kinetic, Coulomb, and exchange energies of the quantum protons.
The classical nuclear-nuclear correction term $ \Delta E^\text{NEO}_{\text{nuc},0}$ corrects the electrostatic interactions between classical nuclei and quantum protons.
This term is expressed as
\begin{equation}
   \Delta E^\text{NEO}_{\text{nuc},0} = \sum_I^\text{nuclei} Z_I\left[v_
   \text{es}^p(\vect r_I) - \sum_{i}\frac{1}{|\vect r_I - \vect r_{i}|}\right]-\frac{1}{2}\sum_{i\neq j}\frac{1}{|\vect r_{i} - \vect r_{j}|},
\end{equation}
where $\vect r_{i},\vect r_{j}$ are the coordinates of the classical point charges in conventional DFT that are treated as quantum protons by NEO, 
which are not necessarily the proton basis function centers. Here $\vect r_I$ is the coordinate of the $I$th classical nucleus.

\subsection{SCF procedure for coupled electron-proton eigenvalue problem}

\begin{figure*}[t]
    \centering
    \begin{tikzpicture}[node distance=1.5cm,
    every node/.style={fill=white}, align=center]
  \node (gs) [c] {Ground state conventional DFT electron density $\rho^e_\text{DFT}$\\+ Initial guess of proton density $\rho^p_\text{initial}$};
  \node (proton_Hamiltonian) [c,below of=gs,yshift=-0cm] {Construct proton Hamiltonian $\hat{H}^p$};
  \node (proton_density) [c,below of=proton_Hamiltonian] {Solve proton KS equation and update proton density $\rho^p$};
  \node (proton_converge) [c,below of=proton_density]  {Proton SCF converged?};
  \node (electron_Hamiltonian) [c,below of=proton_converge]{Update $v_\text{epc}^e, v_\text{es}^p$ in electron Hamiltonian $\hat{H}^e_{\bf k}$};
  \node (electron_density) [c,below of=electron_Hamiltonian]{Solve electron KS equation, update electron density $\rho^e$ \\
  + correct NEO-DFT total energy $E^\text{NEO}_\text{tot}$};
  \node (electron_converge) [c,below of=electron_density]  {Electron SCF converged?};
  \node (end) [c,below of=electron_converge]  {End of NEO-DFT};
   \draw[->] (gs)-- (proton_Hamiltonian);;
   \draw[->] (proton_Hamiltonian) -- (proton_density);
   \draw[->] (proton_density) -- (proton_converge);
   \draw[->] (proton_converge) -- node{No}++(7,0) -- ++(0,3)  -- (proton_Hamiltonian.east);
   \draw[->] (proton_converge) -- node{Yes}(electron_Hamiltonian);
   \draw[->] (electron_Hamiltonian) -- (electron_density);
   \draw[->] (electron_density) -- (electron_converge);
   \draw[->] (electron_converge) -- node{No}++(-7,0) -- ++(0,7.5)  -- (proton_Hamiltonian.west);
   \draw[->] (electron_converge) -- node{Yes}(end);

 \end{tikzpicture}
    \caption{Schematic of NEO-DFT self-consistent loop procedures. The proton SCF loop is nested within the electron SCF loop.} 
    \label{fig:dft_flow}
\end{figure*}
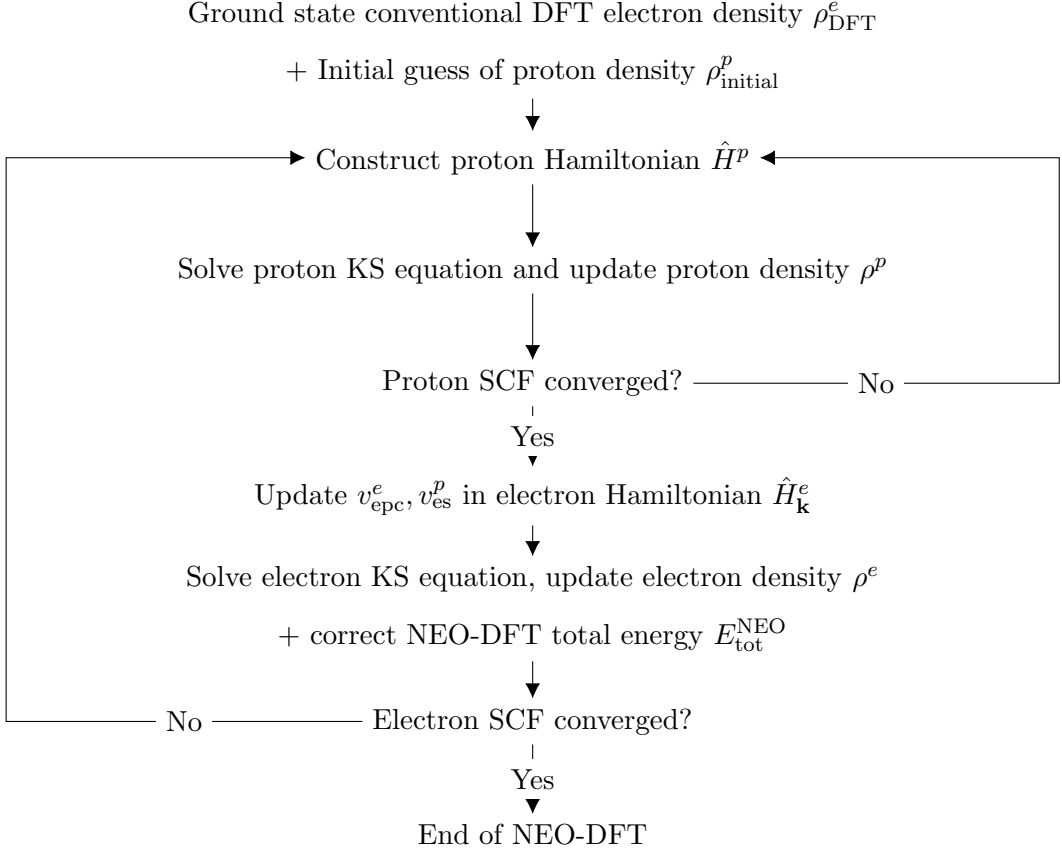

\par 
All necessary ingredients for constructing the electron and proton Hamiltonian are discussed in the preceding sections. 
The generalized eigenvalue problem for the quantum protons is expressed in the GTO basis as $HC=SCE$, with $H,C,E,$ and $S$ denoting the Hamiltonian, eigenvectors, eigenvalues, and  overlap matrix, respectively. To solve this non-orthogonal eigenvalue problem, the Löwdin symmetric orthogonalization method \cite{LOWDIN1970185} is employed, resulting in a standard eigenvalue problem $S^{-\frac{1}{2}}HS^{-\frac{1}{2}}C^{0}=C^{0}E$, where $C^0=S^{\frac{1}{2}}C$.
Eigendecomposition is used to calculate $S^{-\frac{1}{2}}$.
In some cases (e.g., when an even-tempered basis set is used), there is linear dependence in the proton basis, which can lead to a singularity in $S^{-\frac{1}{2}}$. 
To avoid this singularity, eigenvalues of the overlap matrix $S$ that are less than a given threshold (e.g., 1e-6) are removed, and the corresponding eigenspace is neglected in the construction of the inverse matrix. 

Figure \ref{fig:dft_flow} shows the workflow of the SCF procedure for our NEO implementation. 
The proton SCF loop is nested within the electron SCF loop. 
The initial guess of the proton density $\rho^p_\text{initial} = \sum_i|\psi_i^p|^2$ is generated from either a randomly selected Gaussian wave function (i.e., $c_{mi}=\delta_{m,x}$ in Eq. \ref{eq:pwavefunction} with $x$ being the selected GTO basis function) or any preset proton wave function (which could be the converged results obtained by a previous NEO-DFT calculation).
In order to accelerate the convergence of the proton SCF calculation, we adopted the Pulay mixing approach (also know as direct inversion of the iterative subspace,  DIIS).\cite{pulay_improvedscf_1982,kresse_efficiency_1996} 
The proton Hamiltonian is updated according to the DIIS scheme before solving for the proton densities. 
Four convergence criteria are set for the proton SCF iterations:
the maximum matrix element of the proton density matrix difference, 
the change of the sum of the proton eigenvalues,
the integrated proton density difference,
and the maximum matrix element of the error matrix used in DIIS propagation.
The original default convergence criteria in the FHI-aim code are used for the electron SCF iterative procedure.
With this scheme, a similar number of proton SCF cycles (which are nested in each electron SCF cycle) are needed to converge the proton subsystem as compared to the number of electron SCF cycles. For instance, the water-TiO$_2$ system, which is discussed as a validation example in Sec. \ref{sec:validation}, requires about 25 steps to converge the proton SCF cycle per electron SCF step, and it takes 20 electron SCF steps in the outer loop.
This scheme, however, is not the only option, and further improvements are certainly possible in the future.

\section{Validation of Numerical Implementation}
\label{sec:validation}

\begin{table}[b]
    \centering
    \begin{tabularx}{\columnwidth}{XXXX}
        \hline\hline
          && $E_\text{tot}^\text{NEO-DFT}$ & $E_\text{tot}^\text{DFT}$\\
        \hline
         Q-Chem :& No PBC & -2102.5128 & -2103.1777\\
        FHI-aims :& No PBC & -2102.5129 & -2103.1777\\
        FHI-aims :& PBC  10 \AA & -2102.5208 &  -2103.1770\\
         FHI-aims :& PBC 20 \AA & -2102.5161 & -2103.1777\\
         FHI-aims :& PBC  30 \AA & -2102.5129 & -2103.1777\\
          FHI-aims :& PBC 40 \AA & -2102.5129 & -2103.1777\\
        \hline\hline

    \end{tabularx}
    \caption{Total energy (eV) of $\text C_2 \text H_2$ molecule using NEO-DFT ($E_\text{tot}^\text{NEO-DFT}$) and conventional DFT ($E_\text{tot}^\text{DFT}$).
     For the calculations with periodic boundary conditions, a cubic simulation cell is used with the cell length increasing from 10 \AA{} to 40 \AA{}.
    }
    \label{tab:molecule_test}
\end{table}

\par
To the best of our knowledge, there are currently no other periodic DFT implementations of the NEO approach. 
We first validate our implementation against the existing NEO implementation for multicomponent DFT in the Q-Chem code. \cite{epifanovsky2021software} 
We do so by validating our implementation without PBC first and then with PBC using an increasingly larger vacuum region in a cubic simulation cell. 
We consider here an isolated molecule of acetylene ($\text C_2 \text H_2$). 
\footnote{The geometry is optimized with NEO-TDDFT on the first electronic excited state surface. The Cartesian coordinates (in \AA) are (-0.12398,0.672955,0.0) and (0.12398,0.672955,0.0) for the carbon atoms and  (0.69471,1.4301465,0.0) and (-0.69471,-1.4301465,0.0) for the hydrogen atoms.
}
The PB4-D protonic basis set\cite{yu_development_2020} is used for the quantum protons, 
and a 10s10p10d10f even-tempered Gaussian basis set with the exponents ranging from $2\sqrt2$ to 64 is used as the auxiliary basis set for the RI-LVL scheme. \cite{pavosevic_multicomponent_2021}
Convergence tests with respect to the auxiliary basis set can be found in the Supplementary Material.
Both the proton basis functions and the associated RI basis functions are centered at the positions of the classical protons used for the conventional DFT calculations. 
The cc-pVTZ GTO electronic basis set is used here in order to compare to the Q-Chem code, which is based on GTO basis sets. 
We use the epc17-2 electron-proton correlation functional for the NEO-DFT calculations and the B3LYP electronic exchange-correlation functional\cite{becke_densityfunctional_1993,lee_development_1988} for both the conventional DFT and NEO-DFT calculations. 
Table \ref{tab:molecule_test} shows that the non-periodic NEO calculation using our implementation in the FHI-aims code and the existing implementation in the Q-Chem code agree within 0.0001 eV. 
The degenerate occupied proton eigenvalues are calculated to be $-$25.4729 eV using the FHI-aims implementation and $-$25.4787 eV using the Q-Chem implementation, yielding a difference of around 6 meV. 
Given that the FHI-aims and Q-Chem codes utilize very different numerical algorithms for performing electronic structure calculations, the agreement at this level of numerical precision is quite remarkable. 
Table \ref{tab:molecule_test} also shows the convergence of our periodic NEO-DFT calculation as a function of the cubic simulation cell size with the cell length increasing from 10 \AA{} to 40 \AA. 
The periodic NEO-DFT calculation is shown to converge to the non-periodic NEO-DFT calculation when the simulation cell size is sufficiently large so that no artificial interactions among periodic images are present.

\begin{figure}
    \centering
    \includegraphics[width=0.49\textwidth]{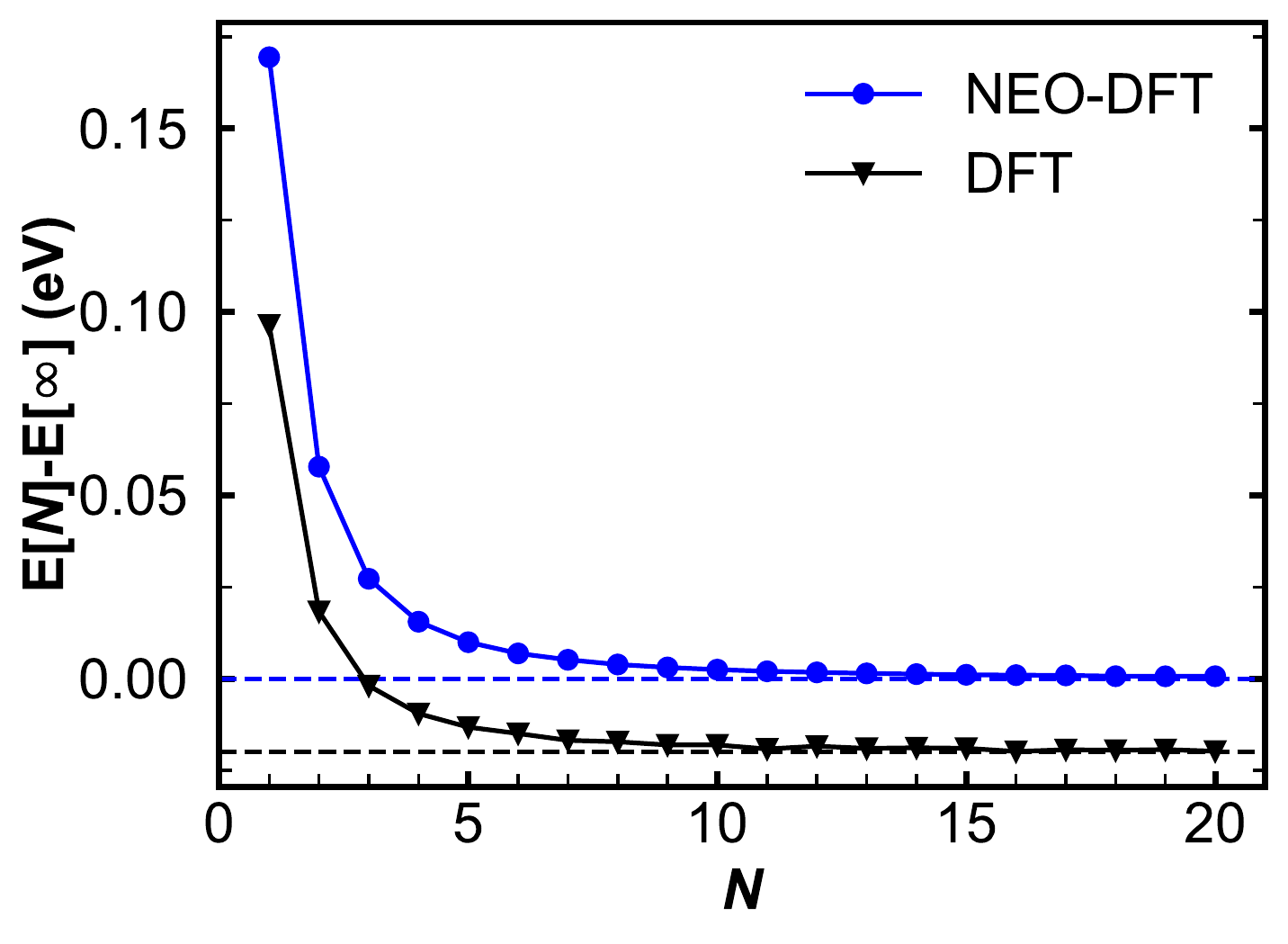}
    \caption{The total energy (in eV) per monomer unit for the  (H$_2$F$_2$)$_N$ oligomer with $N=1 - 20$ using non-periodic calculations (symbols and solid lines), in comparison to that of the  (H$_2$F$_2$)$_{n=\infty}$ polymer using periodic calculations (dashed line).
    Both NEO-DFT and conventional DFT calculations (shifted by -0.02 eV) show convergence with $N \sim 20$.
    For the PBC calculation, the tetragonal unit cell is used with $a=b=50$ \AA, $c=4.33$ \AA, and the converged BZ sampling of
    $10\times1\times1$ k-point mesh is used.
    The oligomer structures are constructed from the primitive unit cell of the infinite polymer chain.}
    \label{fig:molecule_test2}
\end{figure}

\par An additional validation for the periodic NEO implementation is performed by following the procedure described in Ref. \onlinecite{tkatchenko2011unraveling}. 
In this validation test,  the total energy per monomer unit of a one-dimensional chain of (H$_2$F$_2$)$_{n=\infty}$ is computed by using non-periodic calculations on a series of oligomers and by using the periodic calculation on an infinite polymer. 
For sufficiently large $N$, the calculated energy per monomer must be the same such that 
\begin{equation}
    E(\text{monomer}) = \lim_{N \rightarrow \infty} [ E(N+1) - E(N)],
\end{equation}
where $E(\text{monomer})$ is the Brillouin zone converged energy for the unit cell of the polymer using the periodic calculation, and $E(N)$ is the total energy of a finite oligomer chain with $N$ segments using the non-periodic calculation.
The geometry of the zigzag (H$_2$F$_2$)$_n$ chain is taken from Ref. \onlinecite{bezugly2008comparison}.
The default tier1 basis sets and light grid settings in FHI-aims are used for the electronic structure calculations.
The same PB4-D protonic basis set combined with the auxiliary 10s10p10d10f even-tempered Gaussian basis set is used for the RI-LVL scheme.
As shown in Figure \ref{fig:molecule_test2}, the periodic and non-periodic NEO-DFT calculations yield the same total energy per monomer unit within 1 meV when $N$ is sufficiently large. 
The convergence behavior observed here for our NEO-DFT calculation is similar to that observed for the analogous conventional DFT calculation.

\section{Proof-of-Principle Demonstrations}
\label{sec:example}

\par Proof-of-principle calculations of a trans-polyacetylene ($[\text C_2 \text H_2]_n$) chain, a hydrogen boride (HB) sheet, and a titanium oxide-water (TiO$_2$-water) interface are discussed in this section. 
All default parameters in the FHI-aims code are used, and the tier1 basis sets along with light grid settings are used for the  electronic structure calculations in these proof-of-principle demonstrations.
All hydrogen atoms are treated as quantum protons except for the TiO$_2$-water system, where 20 protons in the PBC simulation cell are quantized.
The same 10s10p10d10f even-tempered Gaussian basis set as used in Sec. \ref{sec:validation} is used as the auxiliary basis set for the RI-LVL method.

\subsection{Trans-polyacetylene}
\label{subsec:C2H2}

\begin{table}[b]
    \centering
    \begin{tabularx}{\columnwidth}{XXXXXX}
        \hline\hline
         \multirow{2}{*}{$r_\text{es}^\text{cut}$}& \multicolumn{2}{c}{$[\text C_2 \text H_2]_n$ }& \multicolumn{2}{c}{HB} \\
         \cline{2-5}
         & PB4-D & PB4-F2 & PB4-D & PB4-F2 \\
         \hline
         5 \AA &  -2106.2998& -2106.3200& -1382.1876& -1382.2048\\
         10 \AA & -2106.3245& -2106.3421& -1382.1686& -1382.1893\\
         15 \AA & -2106.3215& -2106.3393& -1382.1697& -1382.1903\\
         20 \AA & -2106.3210& -2106.3388& -1382.1702& -1382.1912\\
         30 \AA & -2106.3208& -2106.3385& -1382.1702& -1382.1912\\
         40 \AA & -2106.3207& -2106.3384& -1382.1701& -1382.1909\\
         \hline
         Ewald & -2106.3208& -2106.3391& -1382.1702& -1382.1915\\
         \hline\hline

    \end{tabularx}
    \caption{Total energy (eV) of $[\text C_2 \text H_2]_n$ and HB sheet calculated using NEO-DFT with two different proton basis sets (PB4-D and PB4-F2) as a function of the radius cutoff ($r_\text{es}^\text{cut}$) values. The values obtained using the standard Ewald summation method are also listed as a reference. 
    }
    \label{tab:rcut_test}
\end{table}

\par 
We will use the simple case of modeling the trans-polyacetylene polymer to first illustrate some key aspects of the periodic implementation of the NEO-DFT method.
The experimental structure of $[\text C_2 \text H_2]_n$ \cite{yannoni1983molecular} is used in the calculations here. 
The B3LYP electronic exchange-correlation functional and the epc17-2 electron-proton correlation functional are used.
First, convergence of the calculation with respect to the  radius cutoff for the proton electrostatic correction potential, $v_\text{es-corr}^p$, is examined (see Sec. \ref{sub:electro}).
For the test case here, 16 k-points are used for sampling the electronic BZ along the periodic axis.
Convergence of the NEO-DFT total energy with the radius cutoff parameter $r_\text{es}^\text{cut}$ ranging from 5 \AA{} to 40 \AA{} is shown in Table \ref{tab:rcut_test} for both the PB4-D and PB4-F2 proton basis sets.\cite{yu_development_2020} 
The total energy converges quickly to 1 meV with the radius cutoff parameter of 20 \AA. 
The fast convergence behavior affirms that the correction potential, $v_\text{es-corr}^p$, can be treated as a short-range potential, especially for cases where protons are well localized.
In terms of computational efficiency, our scheme with the cutoff of 20 $\text{\AA}$ is approximately two times faster than using the conventional Ewald summation method for the trans-polyethylene system. 

\begin{table}[b]
    \centering
    \begin{tabularx}{\columnwidth}{>{\hsize=1.6\hsize\linewidth=\hsize}X>{\hsize=.9\hsize\linewidth=\hsize}X>{\hsize=.9\hsize\linewidth=\hsize}X>{\hsize=.9\hsize\linewidth=\hsize}X>{\hsize=.9\hsize\linewidth=\hsize}X>{\hsize=.9\hsize\linewidth=\hsize}X>{\hsize=.9\hsize\linewidth=\hsize}X}
        \hline\hline
         {No. k points}&
         1& 2& 3& 4& 8& 16\\
        \hline
         {$\Delta E^\text{NEO}/n_p$}&
         0.361& 0.458& 0.463& 0.461& 0.461& 0.461\\
        \hline\hline
         {No. unit cells}&
         1& 2& 3& 4& 8& 16\\
        \hline
         {$\Delta E^\text{NEO}/n_p$}&
         0.361& 0.458& 0.463& 0.461& 0.461& 0.461\\
        \hline\hline
    \end{tabularx}
    \caption{Total energy (eV) difference per quantum proton between NEO-DFT and conventional DFT (i.e.,  the zero-point energy per proton, $\Delta E^\text{NEO}/n_p$) in eV for $[\text C_2 \text H_2]_n$ calculated as a function of the number of k-points used in the BZ sampling and as a function of the number of primitive unit cells in the simulation supercell with the $\Gamma$-point only sampling in the BZ.
    }
    \label{tab:kpoint_test}
\end{table}

\begin{figure*}[t]
    \includegraphics[width=0.99\textwidth]{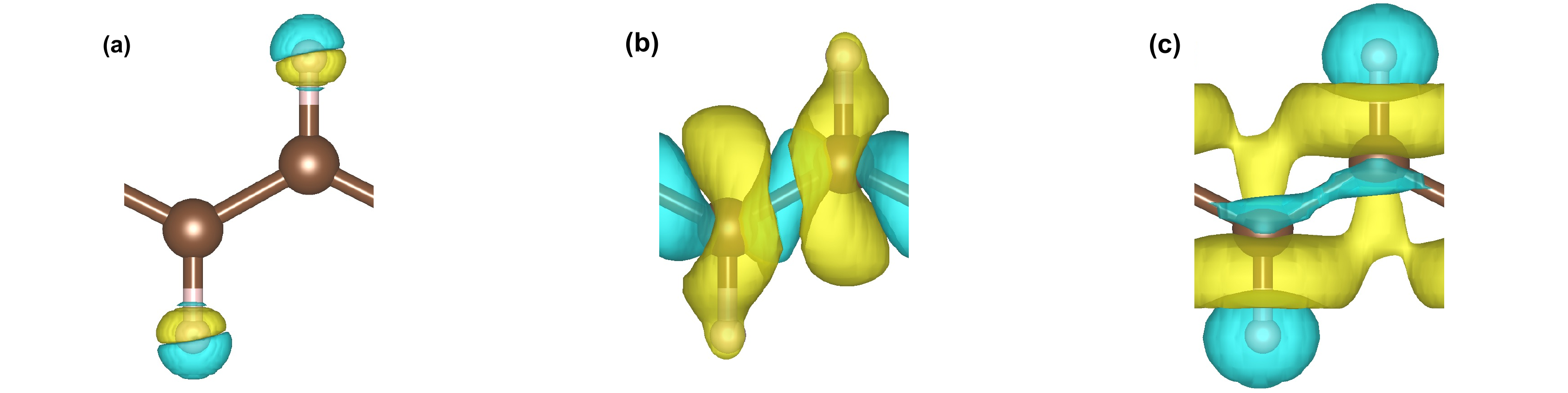}
    
    \caption{Changes in (a) proton density and (b) electron density are shown as the BZ sampling is increased from the $\Gamma$-only approximation to 4$\times$1$\times$1 k-points for electrons in $[\text C_2 \text H_2]_n$. (c) Changes in electron density from conventional DFT to NEO-DFT. Yellow and blue isosurfaces indicate positive and negative changes, respectively. Isosurface values of $8\times 10^{-4}$ \AA$^{-3}$, $2.7\times 10^{-2}$ \AA$^{-3}$ and  $2.7\times 10^{-2}$ \AA$^{-3}$ are used for (a), (b) and (c), respectively. 
    }
    \label{fig:c2h2_density}
\end{figure*}

\par 
We also examine the convergence behavior of our calculation with respect to k-point sampling in the BZ for the electronic structure.
Table \ref{tab:kpoint_test} shows the difference in the total energies between NEO-DFT and conventional DFT per quantum proton ($\Delta E^\text{NEO}/n_p$) for the $[\text C_2 \text H_2]_n$ system. 
$\Delta E^\text{NEO}/n_p$ effectively corresponds to the zero-point energy (ZPE) per proton.  
Two sets of calculations are shown: one for using an  increasingly larger number of k-points with a primitive unit cell of $\text C_2 \text H_2$ unit, and the other for using an increasingly larger supercell containing a larger number of primitive unit cells with the $\Gamma$-point approximation only.
As expected, having a finer sampling of the BZ with a greater number of k-points (O($N_k$) scaling for the computational cost) yields the same ZPE as having more unit cells in the simulation supercell (O($N_{el}^3$) scaling for the computational cost). 

Additionally, these calculations also numerically confirm that the $\Gamma$-point only sampling of BZ integration is sufficient for protons, as discussed in Sec. \ref{sec:method}. Specifically, the proton HOMO eigenvalues of -27.348 eV are the same for both calculations with 16 k-points and 16 unit cells.

Interestingly, the ZPE per proton in $[\text C_2 \text H_2]_n$ is noticeably different from the ZPE in the isolated $\text C_2 \text H_2$ molecule, 0.332 eV. 
The ZPE value of the extended polymer is close to that of the isolated molecule only if the $\Gamma$-point approximation is made (i.e., 0.361 eV vs. 0.332 eV, with minor deviations arising in part from geometrical differences).
When the calculation is converged with respect to the BZ sampling, the ZPE value increases to 0.461 eV, as seen in Table \ref{tab:kpoint_test}.
Figure \ref{fig:c2h2_density} (a) and (b) shows the effects of converging the BZ integration with a greater number of k-points on the proton and electron densities. 
As the BZ sampling is converged, the proton density moves closer to the carbon atoms, while the electron density shifts from the carbon-carbon bonds toward the C-H bonds.
The changes in the proton density are responsible for the increased ZPE value from 0.332 eV to 0.461 eV in transforming  from the isolated $\text C_2 \text H_2$ molecule to the extended $[\text C_2 \text H_2]_n$ polymer system. 
Note that these ZPE values are qualitatively reasonable for the three degrees of freedom associated with each quantum proton. 
We also note that electron-proton correlation cannot be neglected,\cite{doi:10.1021/acs.jctc.1c00454} 
and the ZPE would be as large as 1.203 eV per proton without the electron-proton correlation taken into account.

\begin{figure*}[t]
    \centering
    \includegraphics[width=0.99\textwidth]{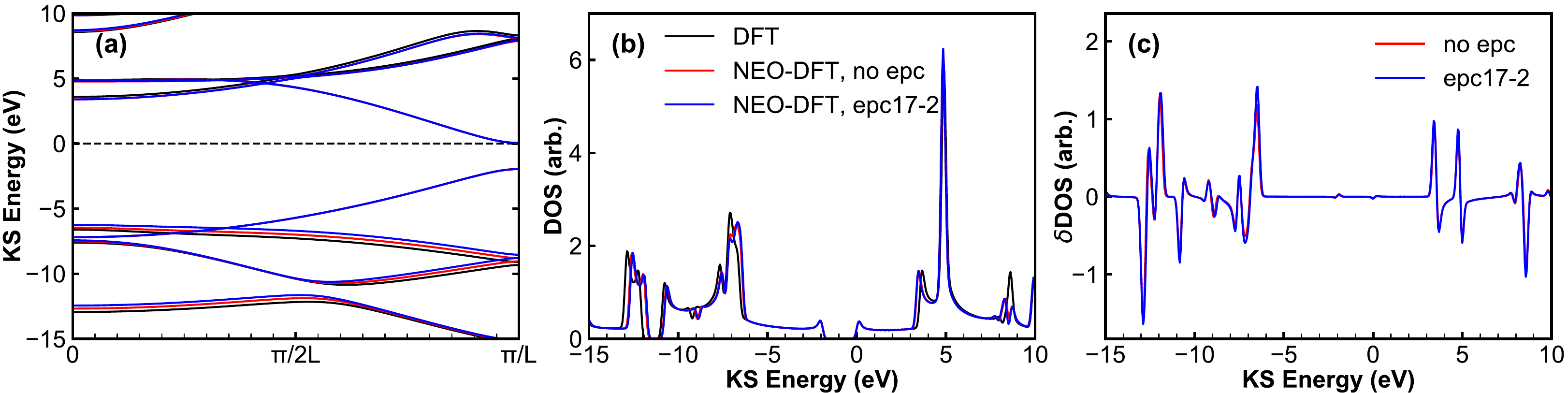}
    \caption{(a) Band structure, (b) DOS, and (c) DOS changes from quantization of protons via the NEO method for the $[\text C_2 \text H_2]_n$ system calculated with conventional DFT (black) and  NEO-DFT with (blue) and without (red) the epc17-2 functional. 100 k-points along the periodic direction are used for the BZ sampling. A broadening factor of 0.1 eV is used in the DOS. The Fermi energy, which is 0.2 eV below the conduction band minimum, is aligned to be 0 eV as a reference.  }
    \label{fig:C2H2}
\end{figure*}

Figure \ref{fig:C2H2} shows the electronic band structure and density of states (DOS) of the $[\text C_2 \text H_2]_n$ system. 
100 k-points along the periodic axis are used for the BZ sampling. The Fermi energy is used as the energy reference (i.e., $E=0$) in all figures. 
With NEO-DFT, the peaks in the DOS are shifted rather substantially at around $-$13 eV and $-$6 eV. 
The band structure also shows corresponding changes in these energy ranges over the entire BZ. 
The magnitudes of the energy shifts are as high as 0.5 eV for some bands, while the energy states near the Fermi energy are not affected by quantization of protons via the NEO method. 
Projection of the DOS on atomic orbitals indicates that the NEO method mainly affects the energy range where hybridization of the 1s electron orbitals of protons and the 2p electron orbitals of carbon atoms form C-H covalent bonds.
Figure \ref{fig:C2H2} also shows the extent to which the DOS and band structure change by including electron-proton correlation through the epc17-2 functional via the multicomponent DFT formulation. 
The electron-proton correlation is found to be essential for the band structure in addition to its crucial role in calculating the ZPE. 
Furthermore, electron-proton correlation exhibits a more profound impact on the proton orbital energies than on  the electron KS eigenvalues.
Including electron-proton correlation using the epc17-2 functional changes the (degenerate) proton HOMO orbital eigenvalue from -27.530 eV to -27.348 eV.


\subsection{Hydrogen boride sheet}

\begin{figure}[t]
    \centering
    \includegraphics[width=0.48\textwidth]{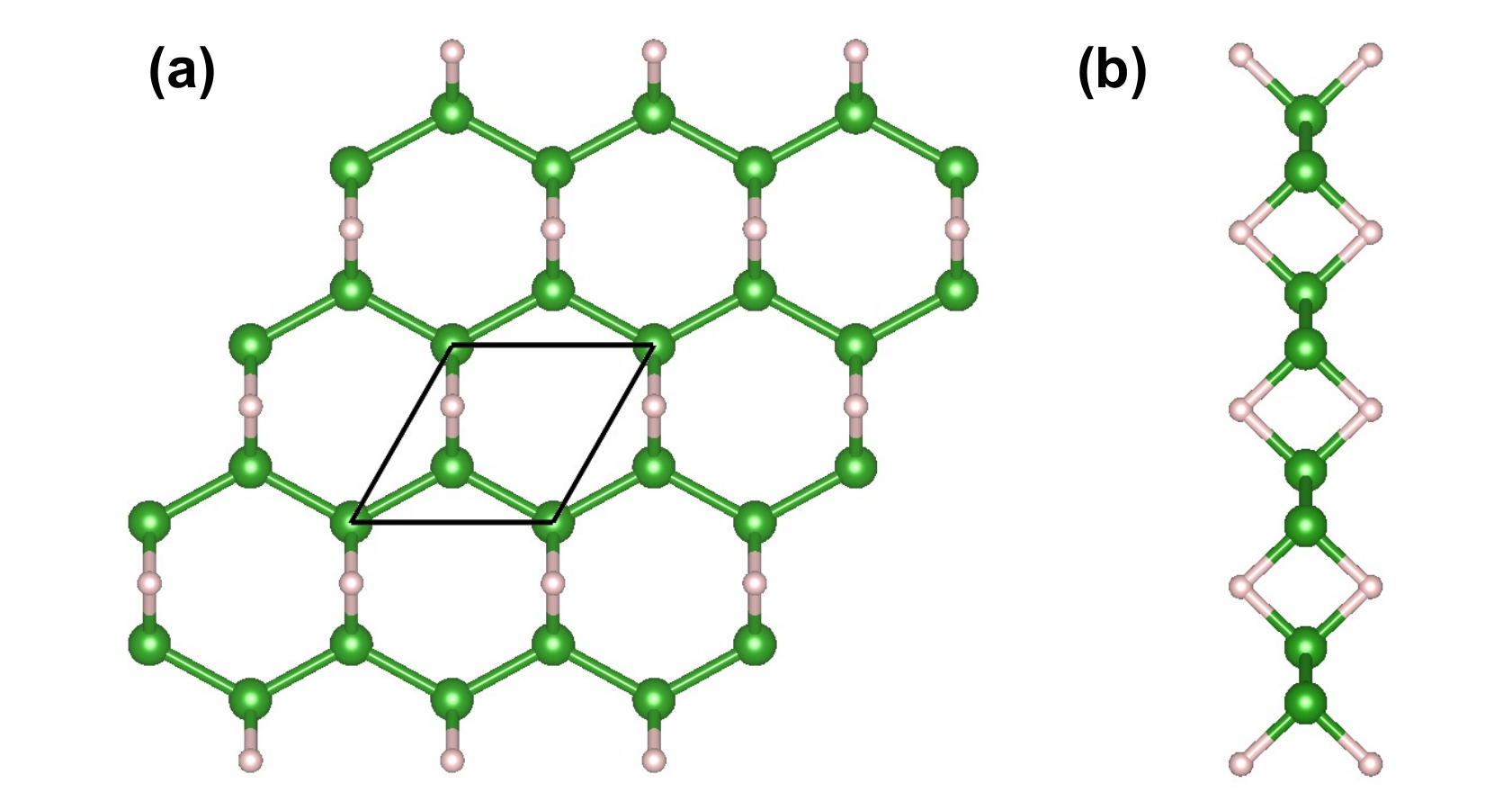}
    \caption{(a) Top view and (b) side view of the HB sheet structure. The primitive unit cell, indicated by the rhombus shape indicated with black lines, includes two boron atoms and two hydrogen atoms. The unit cell is a distorted hexagonal cell with the lattice parameters $a=3.014$ \AA, $b=3.044$ \AA, $c=50$ \AA, $\alpha=\beta=90^{\circ}$, and  $\gamma=60.3172^{\circ}$.  }
    \label{fig:HB_structure}
\end{figure}

\par
\begin{figure*}[t]
    \centering
    \includegraphics[width=0.99\textwidth]{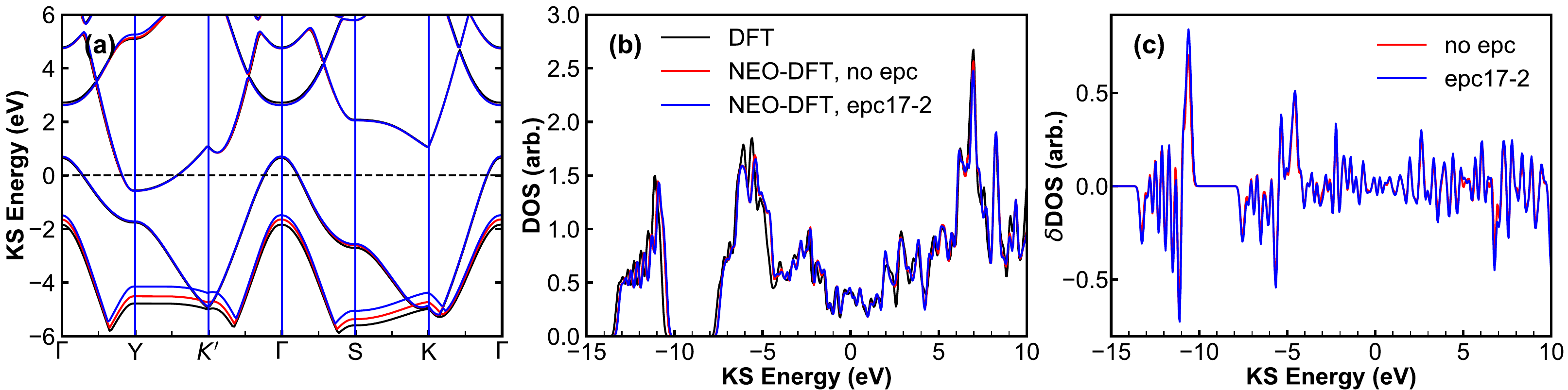}
    \caption{(a) Band structure, (b) DOS, and (c) DOS changes from quantization of protons via the NEO method for the 
    2D HB sheet calculated using conventional DFT (black) and  NEO-DFT with (blue) and without (red) the epc17-2 functional. 
    16$\times$16$\times$1 k-point mesh is used for the BZ sampling. Band structures are calculated along high symmetry k-points of the distorted hexagonal primitive cell. A broadening factor of 0.1 eV is used in the DOS, and the Fermi energy is aligned to be 0 eV as a reference. 
    }
    \label{fig:HB}
\end{figure*}

 Two-dimensional sheets of hydrogen boride (HB) were  recently synthesized, exhibiting some intriguing physical properties for potential applications in hydrogen storage.\cite{nishino_formation_2017,chen_chemical_2020,rojas_chemical_2021} 
The material serves as a good example case in which hydrogen atoms form B-H-B 3-center-2-electron bonds. 
The particular phase of the HB sheet we consider here has the $Cmmm$ space group as shown in Figure \ref{fig:HB_structure}. The DFT-optimized structure is taken from the work by \citeauthor{rojas_chemical_2021}\cite{rojas_chemical_2021} 
The PBE electronic exchange-correlation functional\cite{perdew_generalized_1996} and the epc17-2 electron-proton correlation functional are used in these calculations.
16 $\times$ 16 k-point grids are used for BZ sampling for this metallic system.
The convergence test for the radius cutoff parameter $r_\text{es}^\text{cut}$ is shown in Table \ref{tab:rcut_test}.
A set of high symmetry k-points is used to calculate the band structure as shown in Figure \ref{fig:HB} (a).
The DOS and DOS changes caused by proton quantization with the NEO method for the HB system are shown in Figures \ref{fig:HB} (b) and (c), respectively.
The Fermi energy is set to 0 eV. 
Quantization of the protons alters the DOS and the band structure along a wide energy range. 
The most significant changes are observed for the KS energy range from $-$13 eV to $-$3 eV. 
At the same time, the states near the Fermi level ($-$6 eV $\sim$ $-$2 eV) are also affected by the proton quantization by as much as 0.7 eV.
These calculations also show the importance of quantum protons on the total energy, and the ZPE is calculated to be approximately 0.41 eV per proton.

\subsection{Water at Anatase TiO$_2$ (101) Surface }

\begin{figure*}[t]
    \centering
    \includegraphics[width=0.99\textwidth]{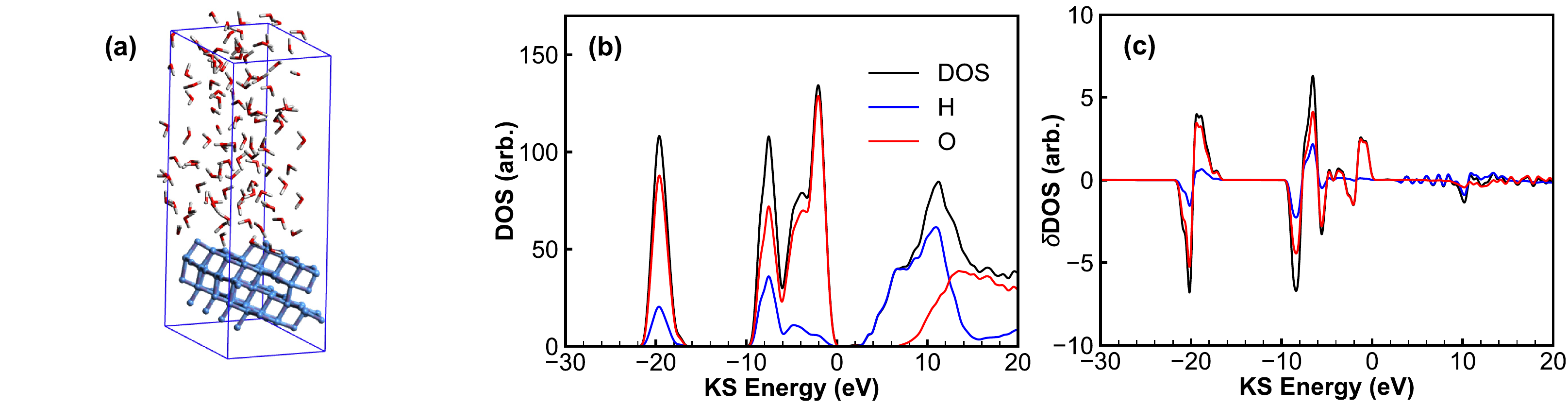}
    \caption{(a) Simulation supercell of TiO$_2$-water interface used in the calculations. (b) PDOS on the water molecules calculated with the NEO-DFT method. The PDOS is further decomposed into contributions from the classical hydrogen (H) and oxygen (O) atomic orbitals of the water molecules. 
    (c) PDOS changes from quantization of protons via the NEO method. 
    A broadening factor of 0.3 eV is used in the DOS. The Fermi level is aligned to be 0 eV as the reference, and the HOMO of water is situated at -0.57 eV.
    }
    \label{fig:TiO2_dos}
\end{figure*}

\par
As a final example of employing the new periodic NEO-DFT method, we  discuss the effects of quantizing protons of water molecules at an anatase TiO$_2$ (101) surface. 
Chemical dynamics of water at this semiconductor surface is of great interest in photo-assisted splitting of water molecules,\cite{nadeem2018water}
and our new periodic NEO method could be of great value in the field.
The TiO$_2$-water interface structure shown in Figure \ref{fig:TiO2_dos}(a) is modeled by taking a snapshot from a first principles molecular dynamics (FPMD) simulation trajectory at an elevated temperature of 330 K, which is necessary for modeling room-temperature water within this level of FPMD simulation. \cite{PhysRevLett.101.017801} 
This proof-of-concept example uses only two layers of anatase TiO$_2$ for studying proton quantization of water at the surface. 
The PBC simulation cell includes 107 water molecules, and the 20 protons closest to the TiO$_2$ top surface are treated as quantum protons in this proof-of-concept demonstration. We also perform another calculation in which 20 protons in the bulk water region, taken here as $18$~\AA{} $<z<20~$\AA{} from the topmost surface layer, are treated quantum mechanically. The electronic structure is modeled using the GGA-PBE exchange-correlation functional with the $\Gamma$-point approximation for BZ integration.
Figure \ref{fig:TiO2_dos} (b) shows the DOS and its projection onto the atomic orbitals of the classical hydrogen and oxygen atoms of the water molecules. 
The Fermi level is set to 0 eV as a reference, and the highest occupied molecular orbital (HOMO) of water is situated at -0.57 eV.
The changes in the DOS and the projected DOS (PDOS) from the conventional DFT to the NEO-DFT calculations are shown in Figure \ref{fig:TiO2_dos} (c).
The quantization of the protons leads to a positive energy shift for the water molecule electronic states below the Fermi level.

The electron-proton correlation is crucial for calculating the total energy in NEO-DFT calculations. 
In terms of the NEO-induced total energy change, for the 20 water molecules closest to the TiO$_2$ surface, the average ZPE (i.e., $\Delta E^\text{NEO}/n_p$) is calculated to be 0.33 eV with the epc-17-2 functional and 1.07 eV without this functional.
For the 20 water molecules in the bulk region away from the surface, the average ZPE is slightly smaller, namely 0.31 eV with the epc17-2 functional and 1.06 eV without it.
The values around 0.3 eV calculated with the epc17-2 functional are reasonable estimates of the ZPE for OH hydrogen-bonded protons and quantitatively agree with a previous study. \cite{doi:10.1021/acs.jctc.1c00454}

\begin{figure*}[t]
    \centering
    \includegraphics[width=0.47\textwidth]{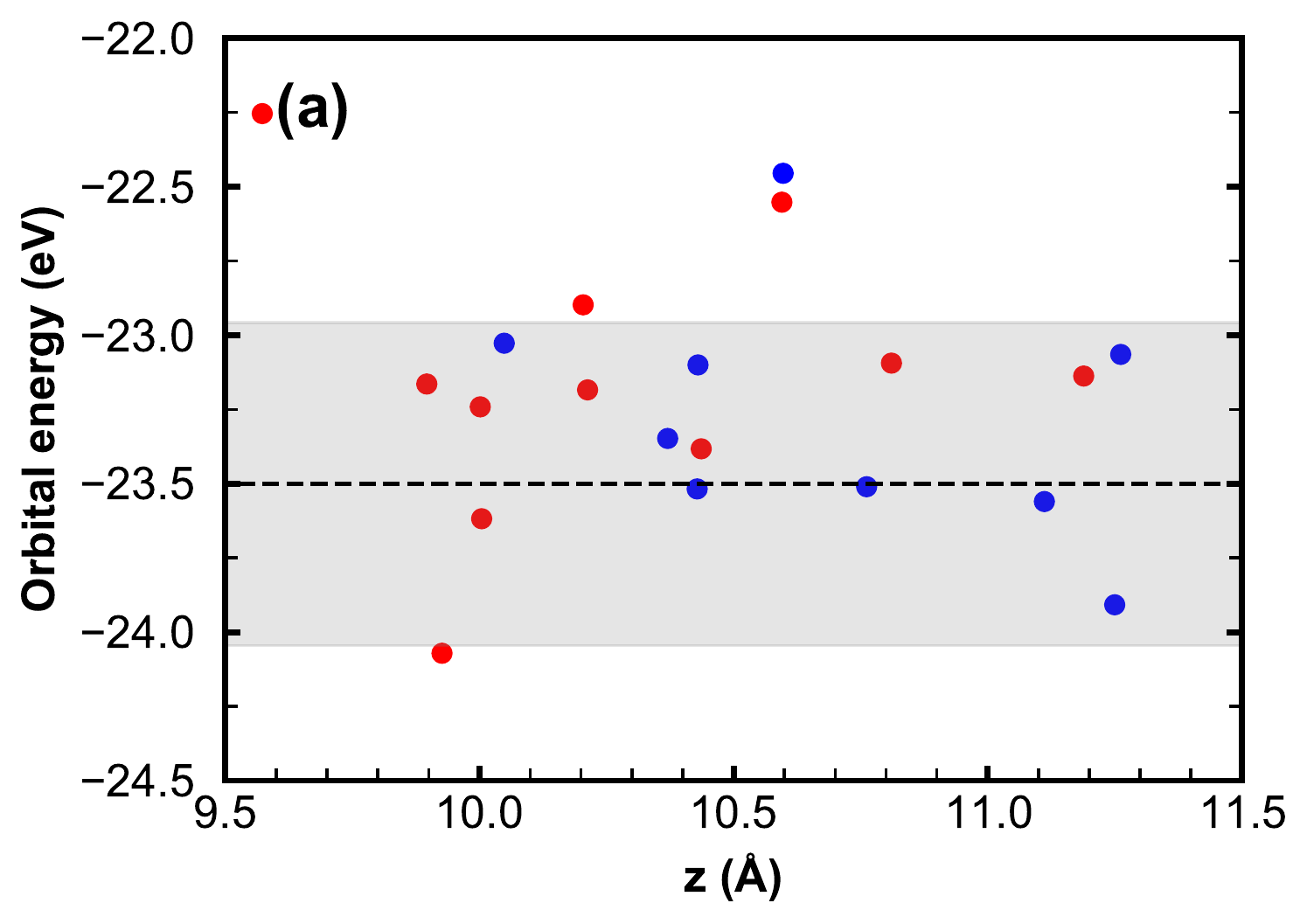}
    \includegraphics[width=0.47\textwidth]{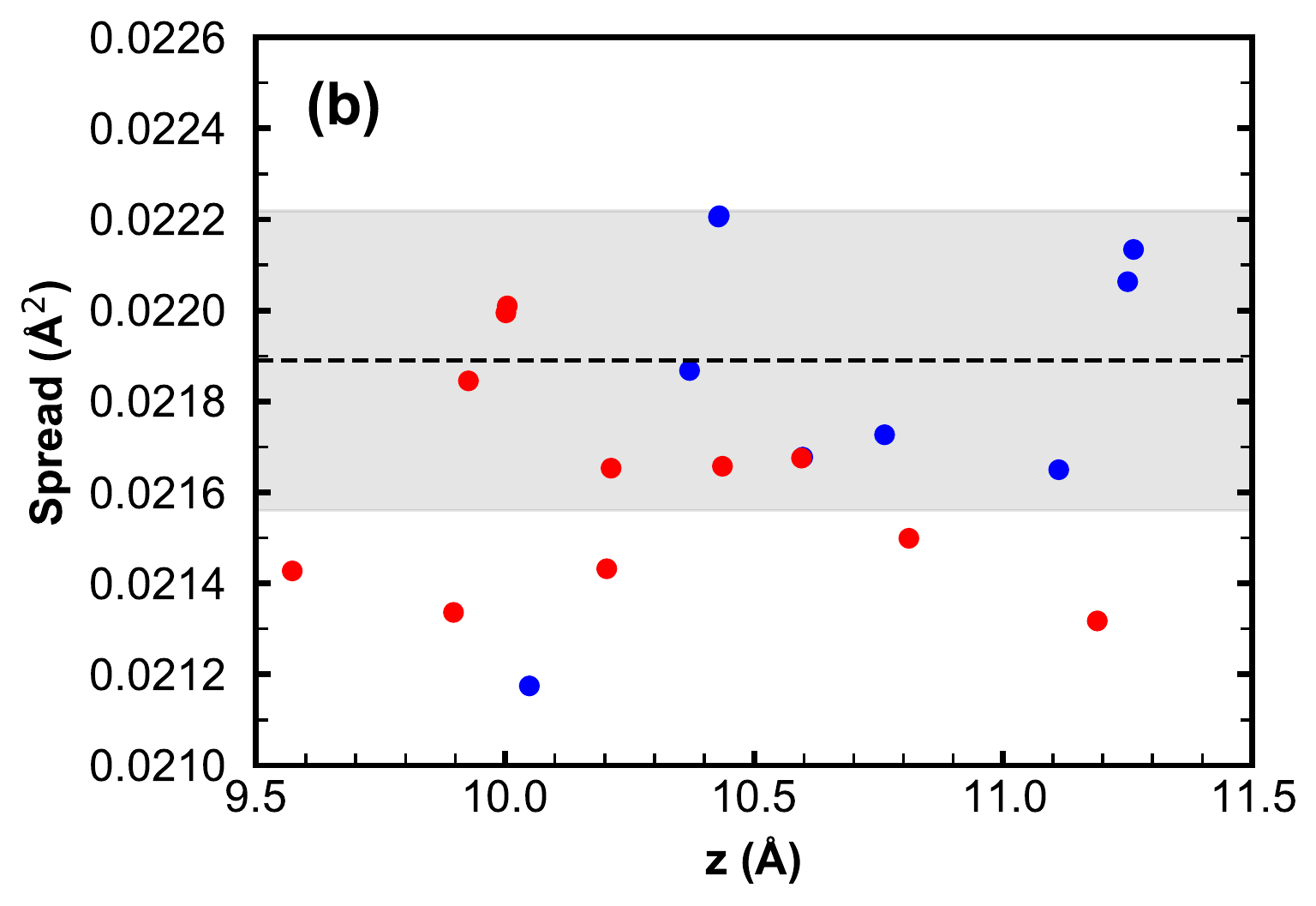}
    \caption{(a) Eigenvalues and (b) spread of localized proton orbitals as a function of their position expectation values in the surface normal direction (i.e., along the z axis). 
    The atoms of the topmost layer of the TiO$_2$ surface are at $z=\sim8.5$ \AA.
    Red and blue colors indicate that the water OH bond for the proton is oriented toward or away from the surface, respectively. 
    The corresponding average values of the protons in the bulk water region ($18$~\AA{} $<z<20$ \AA) are indicated by the dashed lines, and the standard deviations are shown as the gray shaded regions.}
    \label{fig:TiO2_h}
\end{figure*}

\par 
At this water-semiconductor interface, the protons are in distinctively different environments atomistically compared to bulk water, and thus it is instructive to examine how their quantum nature might show some unique characteristics.  
Figure \ref{fig:TiO2_h} shows the expectation values of the proton position operator $\langle\psi^p_n| \hat{\vect r}| \psi^p_n\rangle$ and their spread $\langle\psi^p_n| \hat{\vect r}^2| \psi^p_n\rangle - \langle\psi^p_n| \hat{\vect r}| \psi^p_n\rangle^2$ for the proton orbitals. The position operator in extended systems is defined according to the Resta formula\cite{PhysRevLett.80.1800} in general, but the highly localized nature of the quantum nuclear wave functions allows us to simply employ the position operator through real space integration in practice. The corresponding average values for the protons in the bulk water region are given by the dashed lines, and the grey shaded area indicates the standard deviation. 
The expectation values of the quantum proton positions are located close to the classical proton positions in conventional DFT, and the proton orbitals are highly localized with spread values of $\sim$0.02 \AA$^2$.
The quantum protons at the TiO$_2$ surface are classified based on their oxygen-hydrogen (OH) bond directions with respect to the surface. 
Protons with the OH bond pointing toward the surface are shown as red dots, and protons pointing away from the surface are shown as blue dots.
Water molecules close to the surface tend to orient their OH bonds toward the surface such that the protons form hydrogen bonds with oxygen atoms on the TiO$_2$ surface.
Figure \ref{fig:TiO2_h} shows that protons tend to have higher orbital energies (average values of -23.274 eV for blue dots and -23.145 eV for red dots) and smaller spreads (average values of 0.02190 $\text{\AA}^2$ for blue dots and 0.02162 $\text{\AA}^2$ for red dots) when OH covalent bonds are pointed toward the surface (red dots) due to the electrostatic potential from the TiO$_2$ surface.
Moreover, these properties are likely to depend on the water molecules' local environments, such as hydrogen bonds. 
For example, the particular proton located at $z\sim$9.55 \AA{} shows a rather high orbital energy although its OH bond is still pointed toward the surface at such a short distance.
The higher orbital energy for this particular proton is likely due to the lack of strong hydrogen bonding with a surface oxygen atom. 
Although this simple proof-of-concept demonstration here is not extensive, these observations already show that quantum effects of protons depend on their local chemical environments, and it is an interesting topic for a future study. 
This newly-developed periodic NEO method will help advance our quantum mechanical understanding of such highly complex chemical systems such as water molecules at semiconductor surfaces from first principles.

\section{Conclusion}
\label{sec:conclusion}
In this paper, we report and demonstrate a numerically convenient method for implementing the NEO method in periodic electronic structure codes in the context of multicomponent DFT. 
The NEO method enables us to model selected nuclei quantum mechanically at the same level of theory as the electrons, treating nuclear-electronic correlation in the framework of multicomponent DFT. 
This approach offers a particularly convenient framework for quantizing protons in electronic structure calculations beyond the Born-Oppenheimer ansatz. 
In our implementation, the proton Kohn-Sham equations are solved with information provided by the electron DFT calculation.
Density fitting (i.e., the RI-LVL scheme) is used to capture the electrostatic potential and exchange effects of the quantized protons.
The proton information is passed back to the electron DFT calculation as a correction to the electron Hamiltonian.
The electron and proton Kohn-Sham equations are solved iteratively until a self-consistent solution is obtained.
The modular approach we propose here can also be used with other periodic DFT codes. 
This scheme is demonstrated by implementing it with the FHI-aims code and is tested against an existing implementation in the Q-Chem code for isolated molecular systems. 
We then demonstrate the new periodic NEO-DFT capabilities by presenting three proof-of-principle applications: a polyacetylene polymer, a two-dimensional hydrogen boride sheet, and water molecules at a TiO$_2$ anatase surface.
Our implementation shows that the NEO method has non-trivial effects on the electronic structure of extended systems, and it also enables calculation of nuclear quantum effects (e.g. protonic ZPEs) in condensed matter systems as demonstrated previously for molecular systems.\cite{brorsen_multicomponent_2017}

\par 
As a future direction, we note that electrons and nuclei could be treated at different levels of theory in principle, even combining DFT and wave function theory (WFT), such as DFT in DFT or WFT in DFT,  in the spirit of density functional embedding schemes.\cite{manby2012simple,gomes2012quantum}
The multicomponent framework can also be naturally extended from ground state periodic NEO-DFT calculations to higher-level theories that enable the treatment of excited states. A particularly attractive approach is real-time NEO time-dependent density functional theory (RT-NEO-TDDFT), which has been developed for molecular systems \cite{zhao_real-time_2020} and used to study the nonequilibrium nuclear-electronic dynamics of excited state proton transfer reactions as well as proton vibrational spectra. 
The RT-NEO-TDDFT method also has been combined with Ehrenfest dynamics to describe the nonequilibrium dynamics of all nuclei and electrons. \cite{zhao_nuclearelectronic_2020}
The extension of these approaches to periodic systems will allow the study of a wide range of photocatalytic reactions, such as PCET, in extended condensed matter systems.

\section*{Supplementary Material}

\par See supplementary material for convergence tests of auxiliary basis sets for the RI-LVL method.

\begin{acknowledgments}
This work is based upon work solely supported as part of the Center for Hybrid Approaches in Solar Energy to Liquid Fuels (CHASE), an Energy Innovation Hub funded by the U.S. Department of Energy, Office of Science, Office of Basic Energy Sciences under Award Number DE-SC0021173. This research used resources of the National Energy Research
Scientific Computing Center, a DOE Office of Science User Facility
supported by the Office of Science of the U.S. Department of Energy
under Contract No. DE-AC02-05CH11231 using NERSC award
BES-ERCAP0021125.
\end{acknowledgments}

\appendix

\section{Electrostatic potential of Gaussian multipoles}
\label{app:gau}

\par
Based on the Green's function of the Poisson equation, the electrostatic potential is given by:
\begin{equation}
    U(\vect{r})=\sum_{l=0}^{l_{max}}\sum_{m=-l}^{l}\frac{4\pi}{2l+1}[r^lp^*_{l,m}(r)+\frac{q^*_{l,m}(r)}{r^{l+1}}]Y_{l,m}(\theta,\phi).
    \label{appeq:U}
\end{equation}
Here
\[
    p_{l,m}(r)=\int_r^\infty \int \frac{1}{r'^{l+1}}\rho(r',\theta,\phi)Y_{lm}r'^2d\Omega_{lm}dr'
\]
\[
    q_{l,m}(r)= \int_0^{r} \int r'^{l}\rho(r',\theta,\phi)Y_{lm}r'^2d\Omega_{lm}dr'
\]
The normalized Gaussian multipole density takes the form:
\begin{equation}
    \rho(r',\theta,\phi)=N(\alpha,l)r'^l e^{-\alpha r'^2} Y_{lm}(\theta,\phi)
\end{equation}
where $N(\alpha,l)$ is the normalization coefficient. With the orthonormality of real spherical harmonics 
\begin{equation}
    \int Y_{lm} Y_{l'm'}d\Omega_{lm}=\delta_{ll'}\delta_{mm'},
\end{equation}
the term $p_{l,m}(r)$ can be expressed as
\begin{equation}
\begin{aligned}
    p_{l,m}(r)=&N(\alpha,l) \int_r^\infty  \frac{1}{r'^{l+1}}r'^l e^{-\alpha r'^2}r'^2 dr'=N(\alpha,l) \int_r^\infty   e^{-\alpha r'^2}r' dr'\\
    =&\frac{1}{2}N(\alpha,l) \int_r^\infty   e^{-\alpha r'^2} dr'^2 =\frac{1}{2\alpha}N(\alpha,l) e^{-\alpha r^2} .
    \label{appeq:p}
\end{aligned}
\end{equation}
Similarly, the term $q_{l,m}(r)$ can be expressed as
\begin{equation}
\label{qlm}
    q_{l,m}(r)= \int_0^{r} \int r'^{l}\rho(r',\theta,\phi)r'^2dr'=N(\alpha,l)
    \int_0^{r} r'^{2l+2}e^{-\alpha r'^2}  dr'
\end{equation}
Let us define a functional $Q(n)$ as
\[
\begin{aligned}
    Q(n)\equiv&\int_0^{x} r^{2n+2}e^{-\alpha r^2}  dr \\
    =&\int_0^{x} r^{2n+1}(re^{-\alpha r^2})  dr \\
    =&\int_0^{x} r^{2n+1}(-\frac{1}{2\alpha}e^{-\alpha r^2})  d(-\alpha r^2).
\end{aligned}
\]
Using integration by parts, we find 
\begin{equation}
\begin{aligned}
    Q(n)=&r^{2n+1}(-\frac{1}{2\alpha}e^{-\alpha r^2})|_0^x+\frac{2n+1}{2\alpha} \int_0^{x} r^{2n}(e^{-\alpha r^2})  dr \\
    =&\frac{2n+1}{2\alpha}Q(n-1)-\frac{x^{2n+1}e^{-\alpha x^2}}{2\alpha}
\end{aligned}
\label{appeq:rec_Q}
\end{equation}
where
\[Q(-1)=\int_0^x e^{-\alpha r^2}dr=\frac{\sqrt{\pi}erf(\sqrt{a}r)}{2\sqrt{a}}.\]
Focusing on the terms dependent on the radial $r$ and angular $l$, Equation \ref{appeq:U} can be rewritten as
\begin{equation}
    \label{eq_psi}
    U(\vect r)=\sum_{l=0}^{l_max}\sum_{m=-l}^{l}F_l(r)N(\alpha,l)Y_{l,m}(\theta,\phi),
\end{equation}
where 
\[F_l(r) = \frac{4\pi}{2l+1}[r^lp^*_{l,m}(r)+\frac{q^*_{l,m}(r)}{r^{l+1}}]/N(\alpha,l).\]
Using Eq. \ref{appeq:p} and \ref{appeq:rec_Q} and defining
\begin{align}
    f_l(r) &\equiv [\frac{q^*_{l,m}(r)}{r^{l+1}}]/N(\alpha,l), \label{appeq:rec_q} \\
    g_l(r) &\equiv [r^lp^*_{l,m}(r)]/N(\alpha,l), \label{appeq:rec_p}
\end{align}
the functions $f_l$ and $g_l$ follow the recursion relation 
\begin{equation} f_l = \frac{2l+1}{2\alpha r}f_{l-1} - g_l \end{equation}
\begin{equation} g_l = r g_{l-1}. \label{appeq:rec_g}\end{equation}
Then,
\begin{equation}
\label{appeq:rec_F}
F_l = \frac{4\pi}{2l+1}[f_l + g_l] = \frac{2\pi}{\alpha r}f_{l-1}= \frac{2l-1}{2\alpha r}F_{l-1} - \frac{2\pi}{\alpha r}g_{l-1}.
\end{equation}
Using Eq. \ref{appeq:rec_g} and Eq. \ref{appeq:rec_F}, $F_l$ and $g_l$ can be solved recursively.

\bibliography{NEO}

%

\end{document}